\RequirePackage{pdf14}

\pdfoutput=1
\documentclass[letterpaper]{article}
\usepackage[usenames,dvipsnames,table]{xcolor}
\usepackage{graphicx}
\usepackage{caption}
\usepackage{subcaption}
\usepackage{array,setspace,mathrsfs,amsthm,amsfonts,amsmath,yfonts,dsfont,bbm,colonequals,mathtools}
\usepackage{./aux/jheppub}
\usepackage{afterpage}
\usepackage{xspace}
\usepackage[normalem]{ulem}

\let\oldbfseries=\bfseries
\let\oldmdseries=\mdseries
\let\oldnormalfont=\normalfont
\renewcommand{\bfseries}{\oldbfseries\boldmath}
\renewcommand{\mdseries}{\oldmdseries\unboldmath}
\renewcommand{\normalfont}{\oldnormalfont\unboldmath}

\makeatletter
\newlength{\apb@width}
\newcommand{\autoparbox}[2][c]{\settowidth{\apb@width}{#2}\parbox[#1]{\apb@width}{#2}}

\makeatother

\newcommand{\beqa}{\begin{eqnarray}}
\newcommand{\eeqa}{\end{eqnarray}}
\newcommand{\beq}{\begin{equation}}
\newcommand{\eeq}{\end{equation}}
\newcommand{\lambdan}{\lambda}
\newcommand{\BC}{\textit{BC}}

\newcommand{\wall}{\omega}
\newcommand{\SO}[1]{\mathrm{SO}{(#1)}}

\newmuskip\pFqskip
\mathchardef\pFcomma=\mathcode`,

\newcommand{\cblock}[4]{
	\mathop{#1}\left(\begin{matrix}
		#2 \\
		#3
	\end{matrix}
	; #4 \right)
}

\title{Calogero-Sutherland Approach to Defect Blocks}
\author[1]{Mikhail Isachenkov,}
\author[2]{Pedro Liendo,}
\author[2]{Yannick Linke,}
\author[2]{Volker Schomerus.}

\affiliation[1]{Department of Particle Physics and Astrophysics, Weizmann Institute of
Science, Rehovot 76100, Israel}
\affiliation[2]{DESY Hamburg, Theory Group, Notkestra{\ss}e 85, D-22607 Hamburg, Germany}

\emailAdd{mikhail.isachenkov@weizmann.ac.il}
\emailAdd{pedro.liendo@desy.de}
\emailAdd{yannick.linke@desy.de}
\emailAdd{volker.schomerus@desy.de}

\preprint{DESY 18-070, WIS/08/18-May-DPPA}

\bigskip
\abstract{
Extended objects such as line or surface operators, interfaces or boundaries
play an important role in conformal field theory. Here we propose a systematic
approach to the relevant conformal blocks which are argued to coincide with the
wave functions of an integrable multi-particle Calogero-Sutherland problem. This
generalizes a recent observation in \cite{Isachenkov:2016gim} and makes extensive
mathematical results from the modern theory of multi-variable hypergeometric
functions available for studies of conformal defects. Applications range from
several new relations with scalar four-point blocks to a Euclidean inversion
formula for defect correlators.}


\begin{document}
\setcounter{tocdepth}{2}
\maketitle
\setcounter{page}{1}

\section{Introduction}

Extended objects such as line or surface operators, defects,
interfaces, and boundaries are important probes of the dynamics in
quantum field theory. They give rise to observables that can detect
a wide range of phenomena including phase transitions and non-perturbative
dualities. In two-dimensional conformal field theories they also turned out
to play a vital role for modern formulations of the bootstrap programme.
In fact, in the presence of extended objects the usual crossing symmetry
becomes part of a much larger system of sewing constraints \cite{Cardy:1991tv}.
While initially the two-dimensional bootstrap started from the crossing
symmetry of bulk four-point functions to gradually bootstrap correlators
involving extended objects, better strategies were adopted later which
depart from some of the sewing constraints involving extended objects.
The usual crossing symmetry constraint is then solved at a later stage
to find the bulk spectrum and operator product expansion, see e.g.
\cite{Runkel:2005qw}.

The bootstrap programme, whether in its original formulation \cite{Polyakov:1974gs}, or
in the presence of extended objects, relies on conformal partial wave expansions
\cite{Mack:1973cwx,Ferrara:1973vz} that decompose physical correlation functions into
kinematically determined blocks/partial waves and dynamically determined coefficients.
These conformal blocks for a four-point correlator are functions of two cross-ratios
and the coefficients are those that appear in the operator product expansion of local
fields. Such conformal partial wave expansions thereby separate very neatly the
dynamical meat of a conformal field theory from its kinematical bones.

In order to perform a conformal block expansion one needs a good
understanding of the relevant conformal blocks. While they are in principle
determined by conformal symmetry alone, it is still a highly non-trivial
challenge to identify them in the zoo of special functions.
In the case of scalar four-point functions much progress has been made in the
conformal field theory literature starting with \cite{Dolan:2000ut,Dolan:2003hv,
Dolan:2011dv}. If the dimension $d$ is even, one can actually construct the conformal
blocks from products of two hypergeometric functions each of which depends on one
of the cross-ratios. For more generic dimensions many important properties of
the scalar blocks have been understood, these include their detailed analytical
structure and various series expansions \cite{Pappadopulo:2012jk,Hogervorst:2013sma,
Hogervorst:2013kva,Isachenkov:2017qgn}.

Extended objects give rise to new families of blocks. Previous work on this subject has focused
mostly on local operators in the presence of a defect. This includes correlators and blocks for boundary or defect
conformal field theory \cite{McAvity:1995zd,Billo:2016cpy,Lauria:2017wav,Liendo:2016ymz,Guha:2018snh}, and also
bootstrap studies using a combination of numerical an analytical techniques \cite{Gaiotto:2013nva,
Liendo:2012hy,Gliozzi:2015qsa,Gliozzi:2016cmg,Lemos:2017vnx,Liendo:2018ukf}.\footnote{Related
work includes studies using Mellin space \cite{Rastelli:2017ecj,Goncalves:2018fwx}, and
``alpha space'' \cite{Hogervorst:2017kbj}.} Even in this relatively simple context that involves
no more than two cross-ratios, the relevant conformal blocks were only identified in some special
cases. More general situations, such as e.g. the correlation function of two (Wilson-
or 't Hooft) line operators in a $d$-dimensional conformal field theory,
often possess more than two conformal invariant cross-ratios. Two conformal line operators in a
four-dimensional theory, for example, give rise to three cross-ratios. For a configuration
of a $p$- and a $q$-dimensional object in a $d$-dimensional theory, the number of cross-ratios is
given by $N = \textit{min\/}(d-p,q+2)$ if $p\geq q$ \cite{Gadde:2016fbj}. So clearly, the study of
such defect correlation functions involves new types of special functions which depend on more
than two variables.

In order to explore the features of these new functions, understand
their analytical properties or find useful expansions one could try to
follow the same route that was used for four-point blocks, see e.g.
\cite{Fukuda:2017cup,Kobayashi:2018okw} for some recent work in this direction. It is
the central message of this paper, however, that there is another
route that gives a much more direct access to defect blocks. It relies
on a generalization of an observation in \cite{Isachenkov:2016gim} that
four-point blocks are wave functions of certain integrable two-particle
Hamiltonians of Calogero-Sutherland type \cite{Calogero:1970nt,Sutherland:1971ks}.
The solution theory for this quantum mechanics problem is an important
subject of modern mathematics, starting with the seminal work of
Heckman-Opdam \cite{Heckman:1987}, see \cite{Isachenkov:2017qgn} for
a recent review in the context of conformal blocks. Much of the
development in mathematics is not restricted to the two-particle
case and it has given rise to an extensive branch of the modern
theory of multi-variable hypergeometric functions.

In order to put all this mathematical knowledge to use in the context
of defect blocks, all that is missing is the link between the corresponding
conformal blocks, which depend on $N$ variables, to the wave functions of
an $N$-particle Calogero-Sutherland model. Establishing this link is
the main goal of our paper. Following a general route through
harmonic analysis on the conformal group that was proposed in
\cite{Schomerus:2016epl}, we construct the relevant Calogero-Sutherland
Hamiltonian, i.e.\ we determine the parameters of the potential in
terms of the dimensions $p,q$ of the defects and the dimension $d$. In
the special case of correlations of bulk fields in the presence of
a defect, the parameters also depend on the conformal weights of the
external fields. All these results will be stated in section 3
along with a sketch of the proof.

Calogero-Sutherland models possess a number of fundamental symmetries that
can be composed to produce an exhaustive list of relations between defect
blocks. We will present these as a first application of our approach in
section 3.2. Special attention will be paid to relations involving scalar
four-point blocks for which we produce a complete list that significantly
extends previously known constructions of defect blocks.

As interesting as such relations are, they provide only limited access to
defect blocks. We develop the complete solution theory for defect blocks with
$N=2$ and $N > 2$ cross-ratios in section 4 and 5 by exploiting known
mathematical results on the solutions of Calogero-Sutherland eigenvalue
equations. A lightning review of the mathematical input is included in
section 4, following \cite{Isachenkov:2017qgn}. In particular, we shall
review the concept of Harish-Chandra scattering states, discuss the issue
of series expansions, poles and their residues, as well as global analytical
properties such as cuts and their monodromies. In the final section we put 
all these results together to construct defect conformal partial waves and 
blocks. By definition, the former are linear combinations of conformal 
blocks that are single valued in the Euclidean domain and feature in 
the Euclidean inversion formula. The paper concludes with an outlook and a 
list of important open problems.

\section{Setup and review of previous results}

\newcommand\overmat[2]{%
	\makebox[0pt][l]{$\smash{\overbrace{\phantom{%
					\begin{matrix}#2\end{matrix}}}^{\text{\scriptsize #1}}}$}#2}
\newcommand\undermat[2]{%
	\makebox[0pt][l]{$\smash{\underbrace{\phantom{%
					\begin{matrix}#2\end{matrix}}}_{\text{\scriptsize #1}}}$}#2}

Before we begin discussing our new Calogero-Sutherland approach to defect blocks, we want to summarize
the main results that are present in the existing conformal field theory literature. The setup that has
received most attention involves two bulk fields in the presence of a $p$-dimensional defect. For such
correlators, the conformal blocks are known at least as series expansions \cite{Billo:2016cpy,
Lauria:2017wav} or, more explicitly, through relations with scalar four-point blocks which exist
for some special cases, see subsection 2.3. Results on conformal blocks in the more generic setup when
none of the defects is point-like are particularly scarce, see however \cite{Gadde:2016fbj}, where the
number of independent cross-ratios was counted and a particular set of cross-ratios was constructed.
We shall review some key ingredients from \cite{Gadde:2016fbj} in subsection 2.2. This subsection also
contains a parametrization of defect cross-ratios in terms of new geometric variables that will turn
out to be particularly well adapted to our Calogero-Sutherland models later on.

\subsection{Two-point functions in defect CFT}

In order to describe existing results concerning two bulk fields in the presence of a defect (or boundary),
we briefly review the embedding formalism, which is a standard approach frequently used to study correlators
in conformal field theory. For details on the embedding space formalism see for example \cite{Costa:2011mg}.
The adaptation to the defect setup can be found in \cite{Billo:2016cpy,Gadde:2016fbj}, see also the next
subsection.

Because the Euclidean conformal group in $d$ dimensions is $\SO{1,d+1}$ it is natural to represent its
action linearly on an embedding space $\mathbb{R}^{1,d+1}$. In order to retrieve the usual non-linear action
of the conformal group on the $d$-dimensional Euclidean space we must get rid of the two extra dimensions.
This is done by restricting the coordinates to the projective null cone, i.e. we demand $X^2=0$ for $X \in
\mathbb{R}^{1,d+1}$ and identify $X \sim gX$ for $g \in \mathbb{R}$. It is useful to work in lightcone
coordinates with dot product given by
\begin{equation}
X \cdot Y = (X^+,X^-,X^i) \cdot (Y^+,Y^-,Y^i) = -\frac12(X^+Y^-+X^-Y^+)+X^iY^i \,.
\end{equation}
In other words, points on the physical space $x \in \mathbb{R}^d$ are represented by elements of the
projective lightcone of the embedding space. It is common to use the projective identification $X
\sim gX$ in order to fix a particular section of the cone given by
\beq
X = (1,x^2,x^{\mu})\, .
\eeq
This is called the Poincar\'e section. Note that this section is invariant under $\SO{1,d+1}$ only up
to projective identifications. The point at infinity is lifted to $\Omega = (0,1,0^\mu)$.

Extended operators or defects in conformal field theories do not preserve the $\SO{1,d+1}$ symmetry of
the conformal group. However, if we consider a $p$-dimensional conformal defect its support is left
invariant by the subgroup $\SO{1,p+1} \times \SO{d-p} \subset \SO{1,d+1}$. Indices that transform
non-trivially under the first factor $\SO{1,p+1}$ will be denoted by $A,B, \dots$ while those that
transform non-trivially under the rotation group $\SO{d-p}$ will be denoted by $I,J, \dots$, i.e.\
we split $X\in \mathbb{R}^{1,d+1}$ as
\beq
(X^A) = (X^0, X^1, \ldots, X^{p+1})\, , \qquad (X^I) = (X^{p+2}, \ldots, X^{d+1})\, ,
\eeq
into components along and transverse to the defect. With these introductory remarks on the embedding
space, we are now prepared to discuss two-point functions.

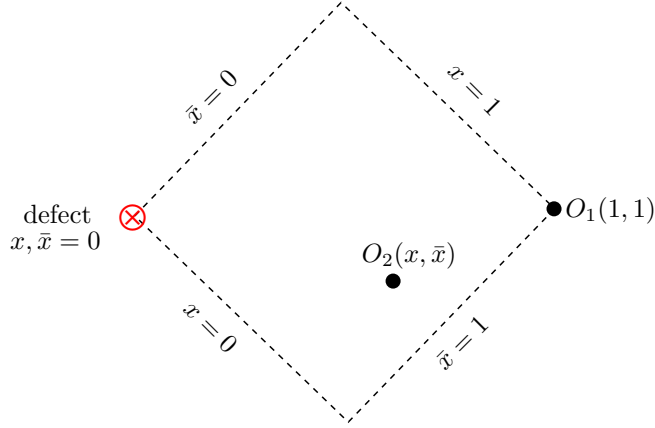
\begin{figure}[htb]
\begin{center}
\includegraphics[scale=0.5]{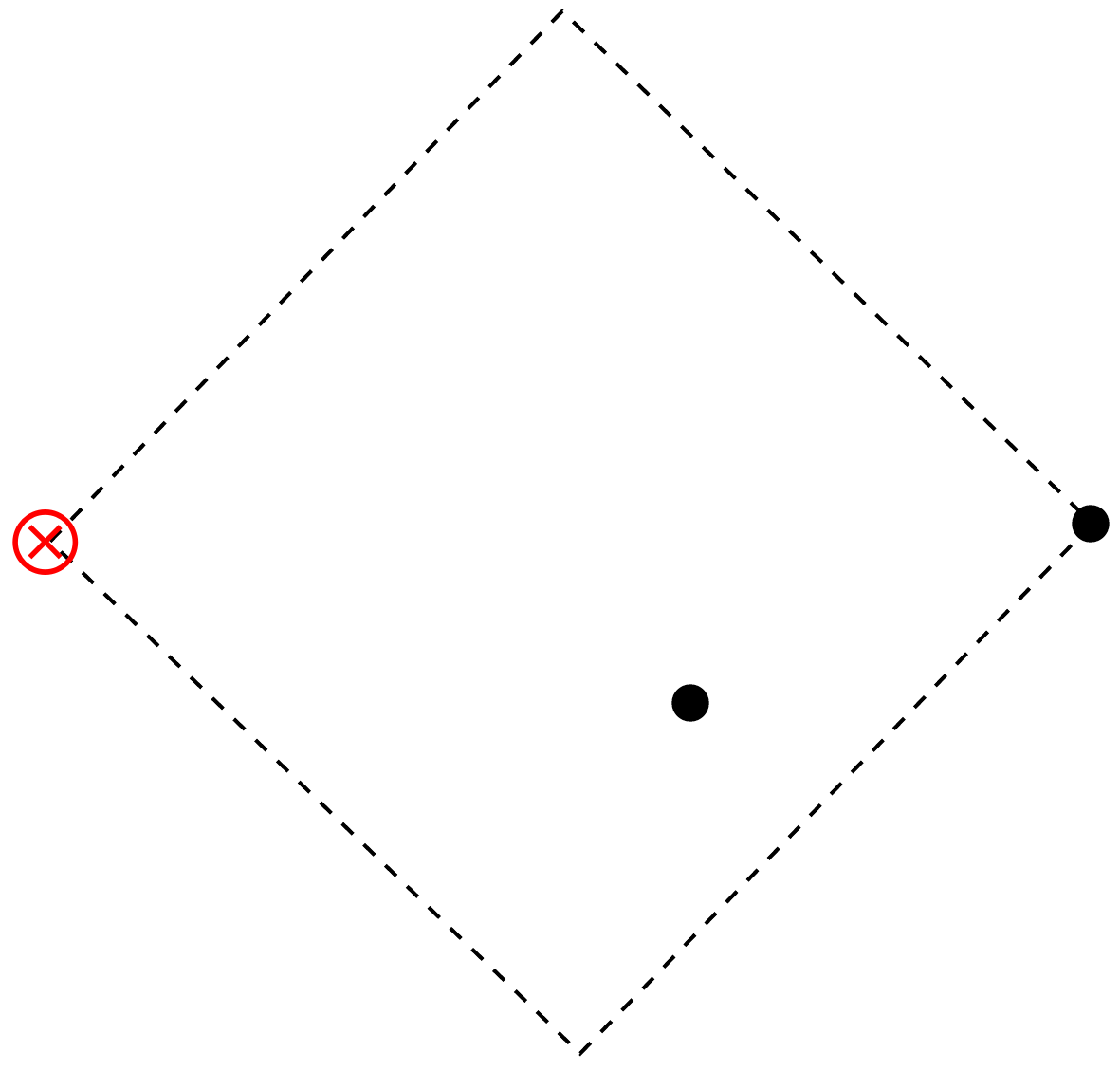}
\put(-210,68){$x,\bar{x}=0$}
\put(-205,78){defect}
\put(-145,115){\rotatebox{45}{$\bar{x}=0$}}
\put(-45,135){\rotatebox{-45}{$x=1$}}
\put(-50,22){\rotatebox{45}{$\bar{x}=1$}}
\put(-145,45){\rotatebox{-45}{$x=0$}}
\put(0,80){$O_1(1,1)$}
\put(-77,63){$O_2(x,\bar{x})$}
\caption{Two-point function configuration in a plane orthogonal to the defect.
The defect is at the origin while the operators $O_1$ and $O_2$ are at points
$(1,1)$ and $(x,\bar{x})$, respectively.}
\label{fig:configuration}
\end{center}
\end{figure}

Let us represent the insertion points of the two bulk fields by $X_1$ and $X_2$. Since the
$p$-dimensional defect splits the $d$-dimensional conformal group into two factors, see previous
paragraph, it is natural to introduce the following product
\beq
X_1 \circ X_2 = X_1^{I} \delta_{IJ} X_2^{J}\, .
\eeq
Here, summation over the transverse indices $I,J = p+2, \dots, d+1$ is understood.
We can now choose two conformal invariants
\beq\label{eq:xxb}
\frac{(1-x)(1-\bar{x})}{(x \bar{x})^{\frac{1}{2}}} = -\frac{2X_1 \cdot X_2}{(X_1 \circ X_1)^{\frac12}(X_2 \circ X_2)^{\frac12}}\, , \qquad
\frac{x+\bar{x}}{2(x \bar{x})^{\frac{1}{2}}} = \frac{X_1 \circ X_2}{(X_1 \circ X_1)^{\frac12}(X_2 \circ X_2)^{\frac12}}\, .
\eeq
The choice of cross-ratios $(x,\bar{x})$ may not appear to be the most natural one at first, but
they turn out have a clean interpretation in terms of  coordinates in a plane orthogonal to the
defect, see figure \ref{fig:configuration}. Conformal symmetry constrains two-point functions to
be of the form
\beq \label{eq:2ptScalars}
\langle O_{1}(X_1) O_{2}(X_2) \rangle^{(p)}_\textit{defect} =
\frac{\mathcal{F}(x, \bar{x})}{(X_1 \circ X_1)^{\tfrac{\Delta_1}{2}}
(X_2 \circ X_2)^{\tfrac{\Delta_2}{2}} }\, ,
\eeq
where the function $\mathcal{F}(x, \bar{x})$ has two conformal block expansions: the bulk channel and the defect
channel to be described below.

\subsubsection{Bulk channel conformal blocks}
The bulk channel expansion is obtained by using the standard operator product expansion
for two local bulk fields before evaluating the one-point functions of the resulting bulk
fields in the background of the defect,
\beq \label{eq:2ptBulkCh}
\mathcal{F}(x, \bar{x}) =\left(\frac{(1-x)(1-\bar{x})}{(x \bar{x})^{\frac{1}{2}}} \right)^{-\frac{\Delta_1+\Delta_2}{2}}
\sum_{k} c_{12k}C^\mathcal{D}_{k}
\cblock{f}{p,a,d}{\Delta_k,\ell_k}{x, \bar{x}}\,,
\eeq
where we made the dependence on the defect dimension $p$, the relevant information about the external scalars $a=(\Delta_2-\Delta_1)/2$, and the dimension $d$ explicit.

The conformal field theory data in this channel corresponds to the bulk three-point coupling $c_{12k}$
multiplied with the coefficients $C^\mathcal{D}_k$ of the one-point function of scalar operators. The general form
of the bulk channel blocks cannot be found in closed-form in the existing literature, see however
\cite{Lauria:2017wav} for efficient power series expansions. For some selected cases the defect block
can be mapped to the conformal blocks for four scalars in standard bulk conformal field theory, see
sections 2.3 and 3.2 below and appendix B. Our results in sections 4-5 generalize these isolated results and thereby
fill an important gap.

\subsubsection{Defect channel conformal blocks}
Local operators in the bulk of a defect conformal field theory may be expanded in terms of operators
that are inserted along the defect. We will denote such operators by $\hat{O}$ and the associated
operator product coefficients for the bulk fields $O_i, i=1,2$ through $b_{i\hat{O}}$. Applying such
a defect expansion to the external operators results in the following conformal block expansion
\beq \label{eq:2ptDefectCh}
\mathcal{F}(x, \bar{x}) = \sum_{k} b_{1 k} b_{2 k} \hat{f}_{\widehat{\Delta}_k,s_k}(x, \bar{x})\, ,
\eeq
where $k$ runs through the set of all intermediate fields $\widehat{O} = \widehat{O}_k$ of weight $\widehat
\Delta_k$ and spin $s_k$. The blocks $\hat{f}(x,\bar{x})$ factorize in terms of the $\SO{d-1,1} \times \SO{d-p}$
symmetry group. This simplifies the analysis significantly and it is possible to write $\hat{f}(x, \bar{x})$ as a
product of hypergeometric functions
$$
\hat{f}_{\widehat{\Delta},s}(x,\bar{x}) = x^{\frac{\widehat{\Delta}-s}{2}}\bar{x}^{\frac{\widehat{\Delta}+s}{2}}
{}_2F_1\left(-s,\frac{d-p}{2}-1,2-\frac{d-p}{2}-s,\frac{x}{\bar{x}}\right) 
{}_2F_1\left(\widehat{\Delta},\frac{p}{2},\widehat{\Delta}+1-\frac{p}{2},x\bar{x}\right)\,.
$$
In the following we shall mostly focus on the bulk channel and its generalizations. A few more
comments on the defect channel and its role in the bootstrap can be found in the concluding section.

\paragraph{Boundary CFT.} As an aside let us comment on the boundary case which is special, since the
transverse space is one-dimensional ($p=d-1$). In this case the two-point function depends only on the
first invariant in eq.\ \eqref{eq:xxb}
\beq
\langle O_{1}(X_1) O_{2}(X_2) \rangle_\textit{BCFT} = \frac{1}
{(X_1 \circ X_1)^{\tfrac{\Delta_1}{2}} (X_2 \circ X_2)^{\tfrac{\Delta_2}{2}}}f\left(\frac{(1-x)
(1-\bar{x})}{(x \bar{x})^{\frac{1}{2}}}\right) \, .
\eeq
The conformal block expansion of this correlator was originally studied in \cite{McAvity:1995zd}, and the
boundary bootstrap was implemented in \cite{Liendo:2012hy,Gliozzi:2015qsa,Gliozzi:2016cmg}.

\subsection{Cross-ratios for two conformal defects}

While some of our new results do concern the configurations considered in the previous
subsection, our approach covers a more general setup involving two defects of dimension
$p$ and $q$, respectively. The first systematic discussion of such defect correlators
can be found in \cite{Gadde:2016fbj}. That paper determined the number $N$ of
cross-ratios and also introduced a particular set of coordinates on the space of
these cross-ratios. Here we shall review the latter before we discuss an alternative,
and more geometric choice of coordinates.

As we have discussed already, a $p+2$-dimensional hyperplane in $\mathbb{R}^{1,d+1}$ with
a time-like direction preserves the subgroup  $\SO{1,p+1} \times \SO{d-p}$ of the conformal
group. Furthermore, it can be shown that the intersection of such a hyperplane with the
Poincar\'e section projects down to a $p$-sphere in $\mathbb{R}^d$ \cite{Gadde:2016fbj}, the
locus of the defect in Euclidean space. Hence, one can parametrize the position of the defect
through $(d-p)$ orthonormal vectors $P_\alpha, \alpha = 1, \dots, d-p$, one for each
transverse direction. In order to do so, we first pick any $p+2$ points $x_k$, $k=1,\dots,p+2$,
on the defect $\mathcal{D}^{(p)} \subset \mathbb{R}^d$ and consider their lift $X_k=(1,x_k^2,x_k)$ to
the Poincar\'e section. This uniquely defines the $(p+2)$-dimensional hyperplane. To select a
set of vectors $P_\alpha$, which are of course not unique, we demand that $X_k \cdot P_\alpha
=0$ and $P_\alpha \cdot P_\beta=\delta_{\alpha\beta}$. Besides conformal transformations, there
also exists an $\mathrm{O}(d-p)$ gauge symmetry which acts on the index $\alpha$, i.e.\ it
transforms the vectors $P_\alpha$ into each other. In order to study the two-point function of
two defect operators $\mathcal{D}^{(p)}(P_\alpha)$ and $\mathcal{D}^{(q)}(Q_\beta)$ that are
inserted along surfaces associated with $P_\alpha$ and $Q_\beta$, respectively, we need to
single out the invariant cross-ratios. Consider the matrix with elements $M_{\alpha\beta} =
P_\alpha \cdot Q_\beta$ of \textit{conformal} invariants. The residual gauge symmetries
$\SO{d-p}$ and  $\SO{d-q}$ which act on the matrix $M$ through left- and right multiplication,
respectively, can be used to diagonalize $M$. The non-trivial eigenvalues provide a complete
set of independent cross-ratios.

To determine their number we need a bit more detail. First, let us consider the case in which
the hyperplanes that are spanned by $P_\alpha$ and $Q_\beta$ have no directions in common.
This requires that $2d-p-q \leq d+2$ or equivalently $d-p \leq q+2$. If we assume $p \geq q$
from now on, the number of cross-ratios is given by $N=d-p$,
\begin{align}
\begin{matrix}
M=\left.\begin{pmatrix}
	* & * & * & * & * \\
	* & * & * & * & * \\
	\undermat{$d-q$}{* & * & * & * & *}
	\end{pmatrix}\right\}\text{\scriptsize $d-p$}\
\mathop{\xrightarrow{\hspace*{2cm}}}\limits^{\SO{d-p}}_{\SO{d-q}}\
\begin{pmatrix}
* & 0 & 0 & 0 & 0 \\
0 & * & 0 & 0 & 0 \\
\undermat{$d-p$}{0 & 0 & *} & 0 & 0
\end{pmatrix}\\
\mathstrut
\end{matrix}
\,.
\end{align}
If $d-p > q+2$, on the other hand, the two hyperplanes spanned by $P_\alpha$ and $Q_\beta$ must
intersect in $d-2-(p+q)$ directions. Hence $d-2-(p+q)$ of the scalar products are invariant and
there are only $d-p-(d-2-(p+q)) = q+2$ nontrivial eigenvalues,
\begin{align}
\begin{matrix}
M=\left.\begin{pmatrix}
	* & * & * & * & * \\
	* & * & * & * & * \\
	\undermat{$d-q$}{* & * & * & * & *}
	\end{pmatrix}\right\}\text{\scriptsize $d-p$}\
\mathop{\xrightarrow{\hspace*{2cm}}}\limits^{\SO{d-p}}_{\SO{d-q}}
\ \begin{pmatrix}
* & 0 & 0 & 0 & 0 \\
0 & * & 0 & 0 & 0 \\
\undermat{$q+2$}{0 & 0} & 1 & 0 & 0
\end{pmatrix}\\
\mathstrut
\end{matrix}
\,.
\end{align}
In total, the number of invariant cross-ratios is therefore $N=\min(d-p,q+2)$. To be
precise, we point out that the full gauge group is actually given by $\mathrm{O}(d-p) \times
\mathrm{O}(d-q)$ and hence the values on the diagonal are only meaningful up to a sign. One
way to construct fully invariant cross-ratios is to consider
\begin{equation} \label{eq:eta}
\eta_a=\mathop{tr} (M M^T)^a \,
\end{equation}
where $a = 1, \dots, N$. This is the set of cross-ratios introduced in \cite{Gadde:2016fbj}.
Here we want to consider a second, alternative set, that is more geometric and also will turn out
later to possess a very simple relation with the coordinates of the Calogero-Sutherland Hamiltonian.

Roughly, our new parameters consist of the ratio $R/r$ of radii of the spherical defects along with
$N-1$ tilting angles $\theta_i$ of the lower $(q-)$dimensional defect in the space that is transverse
to the higher $(p-)$dimensional defect. To be more precise, we place our two spherical defects of
dimensions $p$ and $q$, respectively, such that they are both centered at the origin $\mathbb{R}^d$.
Without restriction we can assume that the $p-$dimensional defect of radius $R$ is immersed in the
subspace spanned by the first $p+1$ basis vectors $e_1,\dots,e_{p+1}$ of the $d$-dimensional
Euclidean space. The radius of the second, $q-$dimensional defect, we denote by $r$. To begin with,
we insert this defect in the subspace spanned by the first $q+1$ basis vectors $e_1,\dots,e_{q+1}$.
Then we tilt the second defect by angles $\theta_1,\dots,\theta_{N-1}$ in the $e_1-e_d,\dots,
e_{N-1}-e_{d+2-N}$ planes, respectively. In other words we act on the locus of the second sphere
with 2-dimensional rotation matrices $R_{(i-1,d+2-i)}(\theta_i)$ in the plane spanned by the basis
vectors $e_{i-1}$ and $e_{d+2-i}$ for $i=1, \dots, N-1$. This gives a well-defined configuration
of defects, because we have $N-1 \leq q+1 \leq p+1 < d+2-N$ for $p \geq q$. With a little bit of
work it is possible to compute the matrix $M$ of scalar products explicitly, see appendix
\ref{apx:coord} for a derivation,
\begin{equation} \label{eq:matrixM}
M = \left(\begin{array}{@{}ccccc|c@{}}
\cosh \vartheta &              &        &                  &   & \\
& \cos\theta_1 &        &                  &   & \\
&              & \ddots &                  &   & \scalebox{1.5}{0} \\
&              &        & \cos\theta_{N-1} &   & \\
&              &        &                  & I & 
\end{array}\right) \quad \textrm{where} \quad \cosh\vartheta = \frac12\left(\frac{r}{R}+\frac{R}{r}\right) \,.
\end{equation}
We shall pick $\vartheta$ to be a positive real number. Using the general prescription
\eqref{eq:eta} the cross-ratios $\eta_a$ that were introduced in \cite{Gadde:2016fbj}
take the form
\begin{equation} \label{eq:etatheta}
\eta_a = \cosh^{2a} \vartheta +  \cos^{2a}\theta_1
+ \dots + \cos^{2a}\theta_{N-1}\,,\quad a=1,\dots,N \,.
\end{equation}
From now on we shall adopt the parameters $\vartheta$ and $\theta_i, i=1, \dots, N-1$  
as the fundamental conformal invariants for $N \geq 3$. While $\vartheta$ can be any
non-negative real number, the variables $\theta_i$ take values in the interval $\theta_i
\in [0,\pi[$.

Let us stress once again, that our geometric parameters $R/r$ and $\theta_i$ represent
just one convenient choice. In the special case with $p = q = d-2$, the variables $\eta_1$
and $\eta_2$ possess a direct geometric interpretation that is based on a slightly different
setup in which one defect is assumed to be flat while the second is kept at finite radius but
displaced and tilted with respect to the first, see \cite{Gadde:2016fbj}. Another important
special case appears for $q=0$, i.e. when two bulk fields are placed in the background of a
defect, which we discussed at length in the previous subsection. In particular, we have
introduced a geometric parametrization of the two cross-ratios, namely through the parameters
$x$ and $\bar{x}$, see eq.\ \eqref{eq:xxb}. It is not too difficult to work out, see appendix
\ref{apx:coord}, that these are related to the parameters $\vartheta$ and $\theta\equiv\theta_1$
through
\begin{equation} \label{eq:xxbtheta}
x = \tanh^{-2} \frac{\vartheta+i\theta}{2} \,,\quad
\bar{x} = \tanh^{-2} \frac{\vartheta-i\theta}{2} \,.
\end{equation}
We will use the coordinates $x$, $\bar{x}$ as the fundamental conformal invariants for $N = 2$.
Eq.\ \eqref{eq:xxbtheta} also shows that the variables $\vartheta$ and $\theta_i$ generalize  
the radial coordinates that were introduced for $N=2$ in \cite{Lauria:2017wav}.  

\subsection{Defect partial wave expansion and blocks}

After having identified the variables, we can write down the two-point function of defects
$\mathcal{D}^{(p)}(P_\alpha)$ and $\mathcal{D}^{(q)}(Q_\beta)$, i.e. generalize eqs.\
\eqref{eq:2ptScalars} and \eqref{eq:2ptBulkCh} to an arbitrary pair of defects. Conformal
invariance restricts its form to be
\begin{equation} \label{eq:CPWexp}
\langle \mathcal{D}^{(p)}(P_\alpha) \mathcal{D}^{(q)}(Q_\beta) \rangle = \sum_{k}
C^{\mathcal{D}^{(p)}}_k C^{\mathcal{D}^{(q)}}_k \cblock{f_D}{p,q,d}{\Delta_k,\ell_k}{\vartheta,\theta_i} \,,
\end{equation}
where the spin $\ell$ is labeled by a set of even integers $\ell = (l_1, \dots , l_{N-1})$ with $l_1 \geq \dots \geq l_{N-1} \geq 0$ and the defect blocks $f_D$ are normalized such that
\begin{equation}\label{eq:defblocknorm}
\cblock{f_D}{p,q,d}{\Delta,\ell}{\vartheta,\theta_i} \stackrel{\vartheta\rightarrow\infty}{\rightarrow} 4^\Delta e^{-\Delta\vartheta} \prod_{i=1}^{N-1} (-2\cos\theta_i)^{l_i} \,,
\end{equation}
so that $C^{\mathcal{D}^{(p)}}_k$ are the coefficient in the \textit{defect expansion}
of the defect in terms of local bulk operators
\begin{equation}
	\mathcal{D}(P_\alpha) = \sum_{\Phi} C^{\mathcal{D}}_\Phi D_{\Delta_\Phi}
(P_\alpha, X, \partial_X) \Phi(X) \,.
\end{equation}
The partial wave expansion \eqref{eq:CPWexp} separates these dynamical data from the kinematical
skeleton of the correlation function. The latter enters through the conformal blocks $f_D(\vartheta,\theta_i)$
which are the main objects of interests for the present work. As we mentioned before, these blocks are known
in a few examples where they can be related to the blocks of four scalar bulk fields.

The first example we want to discuss here is taken from \cite{Liendo:2016ymz}. It applies to the case in
which two bulk fields in $d=4$ dimensions are inserted into the background of a line defect, i.e. $p=1$
and $q=0$. In order to relate the defect block $f(x,\bar{x})$ to the blocks $g(\gamma,\bar{\gamma})$
of four scalar fields, let us consider the following change of coordinates
\begin{equation}\label{eq:defectcoordz}
\gamma = \left(\frac{1-x}{1+x}\right)^2 \,,\quad \bar{\gamma} = \left(\frac{1-\bar{x}}{1+\bar{x}}\right)^2 \,.
\end{equation}
which maps the Euclidean region of the defect coordinates $x,\bar x$ to the Euclidean region of the
four-point cross-ratios $\gamma$, $\bar{\gamma}$. Given this change the following identity
holds \cite{Liendo:2016ymz}
\beq \label{eq:secondrel}
\cblock{f}{1,0,4}{\Delta, \ell}{x,\bar x} \propto (\gamma\bar{\gamma})^{-\frac14}
\cblock{g}{\frac14,-\frac14,3}{\frac{\Delta+1}{2}, \frac{\ell}{2}}{\gamma, \bar{\gamma}} \,.
\eeq
The lower indices on the block $g$ refer to the conformal weight and spin of the intermediate field.
The upper indices $(a,b,d) = (1/4,-1/4,3)$ contain the relevant information about the external scalars,
i.e.\ the parameters $a =(\Delta_2-\Delta_1)/2, b = (\Delta_3-\Delta_4)/2$ and the dimension $d$.
Note that the four-point block on the right hand side is the one with $\Delta_1 - \Delta_2 = \Delta_3 - \Delta_4 = - 1/2$
and dimension $d=3$ even though the original defect setup is in $d=4$ dimensions and involves two
bulk fields of the same weight.

A second example for a relation between defects and scalar four-point was pointed out in \cite{Gadde:2016fbj}.
Conformal blocks for the two-point function of defects of dimension $p=q=d-2$ can be mapped to the four-point
function of scalars with the following relation between the different variables
\beq \label{eq:rel3}
\eta_1 = \frac{2(1+v)}{u}\, , \quad \eta_2 = \frac{2(1+6v+v^2)}{u^2}
\eeq
where $u$ and $v$ are related to the usual cross-ratios $z$ and $\bar{z}$ as $u = z\bar{z}$ and
$v = (1-z)(1-\bar{z})$. The relation between $\eta_a$ and $\theta_1,\vartheta$ is given in eq.\
\eqref{eq:etatheta}. With this change of variables the relation of \cite{Gadde:2016fbj} reads
\beq \label{eq:thirdrel}
\cblock{f_D}{d-2,d-2,d}{\Delta, \ell}{\vartheta,\theta_1} = \cblock{g}{0,0,d}
{\Delta, \ell}{z,\bar{z}} \,.
\eeq
As in the previous example, the Euclidean region of the defect block is mapped to a pair of
complex conjugate variables $z, \bar z$ and hence to the Euclidean region of the four-point
blocks. The scalar block on the right hand side is the one with $a=b=0$ and the same
dimension $d$ as on the left hand side.

Another relation between blocks was proposed in \cite{Billo:2016cpy} (chronologically this was the first such relation found). These authors considered
two bulk fields, i.e. $q=0$,  in the presence of a defect of dimension $p = d-2$ and found the
following relation between the corresponding defect blocks in the bulk channel with four-point blocks:
\beq \label{eq:firstrel}
\cblock{f}{d-2,0,d}{\Delta, \ell}{x,\bar{x}} \sim \cblock{g}{0,0,d}{\Delta, \ell}{1-x,1-\frac{1}{\bar{x}}} \,.
\eeq
Here we should point out however, that this relation does not map the Euclidean region of
the defect block to the Euclidean region of the scalar four-point block. In fact, it maps two Lorentzian regions into each other, see also \cite{Lauria:2017wav}. Hence, any relation of the form
\eqref{eq:firstrel} involves an analytic continuation. Since the blocks possess branch cuts, this continuation requires additional
choices. In this case, the lightcone OPE implies that the ambiguity is just a global phase, and indeed \eqref{eq:firstrel} gives the
correct defect block.\footnote{We thank Marco Meineri for discussions and clarifications about this point.}
Nevertheless, the r.h.s.\ of \eqref{eq:firstrel} is not a Euclidean four-point block, but the analytic continuation of such, this is why we put a $\sim$ instead of an equality. We will come back to this issue in section 5.

As we will see, the technology presented in the next section will explain all these relations
and vastly generalize them, through a (re-)interpretation as symmetries of Calogero-Sutherland
models.

\section{Calogero-Sutherland model for Casimir equations}

In this section we want to describe a fully systematic framework for the Casimir
equations of conformal blocks for correlation functions of two defects. Rather
than working with the popular embedding space, we shall realize all blocks as
functions on the conformal group itself. If the latter is equipped with an
appropriate set of coordinates, the Casimir equations assume a universal form.
In fact, they can be phrased as an eigenvalue problem for an $N$-particle
Calogero-Sutherland system. We will review the result in the first subsection,
discuss some immediate consequences of the equations and their symmetries in the
second and sketch the derivation of our results in the third.

\subsection{Calogero-Sutherland models for defects}

We will show below that the Casimir equations for conformal blocks of two defects
can be restated as an eigenvalue problem for the Calogero-Sutherland Hamiltonian
of the form
\begin{align}\label{eq:HCS}
H_\textrm{CS} &= -\sum_{i=1}^{N} \frac{\partial^2}{\partial \tau_i^2} +
\frac{k_3(k_3-1)}{2}\sum_{i<j}^N\left[ \sinh^{-2}\left(\frac{\tau_i+\tau_j}{2}\right)+
\sinh^{-2}\left(\frac{\tau_i-\tau_j}{2}\right) \right]
\notag\\[2mm]
&\qquad
+ \sum_{i=1}^{N}\left[ k_2(k_2-1)\sinh^{-2}\left(\tau_i\right) +
\frac{k_1(k_1+2k_2-1)}{4}\sinh^{-2}\left(\frac{\tau_i}{2}\right) \right] \ .
\end{align}
The coupling constants $k_i, i= 1,2,3$ that appear in the potential are referred to
as multiplicities in the mathematical literature. In principle, these can assume
complex values though we will mostly be interested in cases in which they are real.
The coordinates $\tau_i$ may also be complex in general. Later we will describe their
values in more detail. The case $N=1$ is a bit special since it involves only two
coupling constants.

The Calogero-Sutherland Hamiltonian possesses two different interpretations. We can
think of it as describing a system of $N$ interacting particles that move on a
one-dimensional half-line with external potential. The external potential is given
by the terms in the second line of eq.\ \eqref{eq:HCS}. These terms contains two of the
three coupling constants, namely $k_1$ and $k_2$. The interaction terms, on the other
hand, involve the third coupling constant $k_3$. Alternatively, we can also think
of a scattering problem for a single particle in an $N-$dimensional space. We will
mostly adopt the second view below.

Let us note that the multiplicities are not defined uniquely, i.e. different choices
of the multiplicities $k_i$ can give rise to identical Casimir equations. This is
partly due to the fact that the multiplicities appear quadratically in the potential.
In addition, one may show that a simultaneous shift of all coordinates $\tau_i \rightarrow
\tau_i + i\pi$ for $i=1, \dots, N$ leads to a Calogero-Sutherland Hamiltonian of the form
\eqref{eq:HCS} with different multiplicities. The complete list of symmetries is given
in table \ref{tab:sym}. Later we see that these innocent looking replacements have
remarkable consequences, since they produce non-trivial relations between the blocks of
various (defect) configurations.
\begin{table}[h!]
  \centering
  \caption{Symmetries of the Calogero-Sutherland model for generic values of the
  multiplicities. The last symmetry also involves a shift $ \tau'_i = \tau_i \pm
  i \pi$ of the coordinates.}
  \label{tab:sym}
  \begin{tabular}{l|c|c|c}
  & $k'_1$ & $k'_2$ & $k'_3$ \\[2mm] \hline & & & \\[-3mm]
  $\varrho_1$ & $1-k_1-2k_2 $ & $  k_2 $ & $  k_3 $ \label{eq:k1} \\[2mm]
  $ \varrho_2$  & $ -k_1 $ & $  1-k_2 $ & $ k_3 $ \label{eq:k2} \\[2mm]
  $\varrho_3$ & $ k_1 $ & $  k_2 $ & $  1-k_3 $ \label{eq:k3} \\[2mm]
  $ \tilde \varrho$ & $ k_1 $ & $ 1-k_1-k_2 $ & $ k_3 $ \label{eq:marco}
  \end{tabular}
\end{table}
\medskip

Let us now describe the main new results of this work. The first case to look at is
the case of two defects of dimension $p \geq q$ with $q \neq 0$. The corresponding
Casimir equation for conformal blocks is an eigenvalue equation for the operator
\begin{equation} \label{eq:Caspq}
L^2 = H_{CS} + \epsilon_0 \,,\quad \epsilon_0 = \frac{N}{8}\left(\frac{d(d+2)}{2} - N(d+1) + \frac{2N^2+1}{3}\right)
\end{equation}
with the following choice of parameters
\begin{align} \label{eq:Caspqpar}
N&=\min(d-p,q+2) \,,\quad k_1=\frac{d}{2}-(p-q)-N+1 \,,\quad k_2=\frac{p-q}{2} \,,
\quad k_3=\frac12 \,.
\end{align}
Let us note that in a representation of spin $\ell$ and weight $\Delta$, the operator
$L^2$ assumes the value
\begin{equation}
	C_{\Delta,J} = \Delta(\Delta-d) + \sum_{i=1}^{N-1} l_i(l_i+d-2i) \,,
\end{equation}
where the spin $\ell$ is labeled by a set of even integers $\ell = (l_1, \dots , l_{N-1})$ with $l_1 \geq \dots \geq l_{N-1} \geq 0$.
The wave function $\psi(\tau)$ is given by the Schr\"odinger-like equation
\begin{equation} \label{eq:schrodinger}
	H_{CS}\psi_\epsilon(\tau) = \epsilon \psi_\epsilon(\tau)
\end{equation}
and is related to the conformal block by\footnote{We postpone the normalization to section 5.2.}
\begin{equation}\label{eq:cb_wavefct}
	\cblock{f_D}{p,q,d}{\Delta,\ell}{\tau} =  2^{2\Delta-\frac12N(d-N+1)} \omega(\tau)\psi_\epsilon(\tau) \,,\quad \epsilon =
-\frac14 C_{\Delta,\ell} - \epsilon_0 \,,
\end{equation}
where the ``gauge transformation'' $\omega(\tau)$ is given by
\begin{align}\label{eq:gauge1}
\omega (\tau) &= \prod_{i=1}^{N}\sinh^{N-\frac{d}{2}+\frac{p-q}{2}-1}\left(\frac{\tau_i}{2}\right)
\cosh^{-\frac{p-q}{2}}\left(\frac{\tau_i}{2}\right) \prod_{i < j}\sinh^{-\frac12}\left(\frac{\tau_i \pm \tau_j}{2}\right).
\end{align}
Here and throughout the entire text below we use the shorthand
\begin{align}
\sinh\left(\frac{x\pm y}{2}\right) =
 \sinh\left(\frac{x+y}{2}\right)
\sinh\left(\frac{x-y}{2}\right) \,.
\end{align}
Equation \eqref{eq:schrodinger} is to be considered on a subspace of the semi-infinite
hypercuboid $A^E_N$ that is parametrized by the coordinates
\begin{align} \label{eq:xrange}
\tau_1 = 2\vartheta =  2\log\frac{R}{r}\in [0,\infty) \,,\quad \tau_{j+1} &= 2i\theta_{j}\in i[0,2\pi] \,,
\end{align}
for $j=1, \dots, N-1$. We shall discuss the domain in much more detail in section 5.1. Of 
course, the choice of multiplicities $k_i$ is not unique since we can apply
any of the transformations listed in table \ref{tab:sym}. We will discuss the consequences
in the next subsection.
\medskip

If $q=0$ while $0< p \leq d-2$, the setup describes two scalar bulk fields in the
presence of a $p$-dimensional defect of co-dimension greater or equal to two. In
this case, the conformal Casimir operator takes the form
\begin{equation} \label{eq:Casp0}
L^2 = H_{CS} + \epsilon_0 \,,\quad \epsilon_0 = \frac{d^2-2d+2}{8}
\end{equation}
with parameters
\begin{align} \label{eq:Casp0par}
N&=2 \,,\quad k_1=\frac{d}{2}-p-1 \,,\quad k_2=\frac{p}{2} \,,\quad k_3=\frac12 + a \,.
\end{align}
Here, the parameter $a$ is related to the conformal weights $\Delta_1$ and $\Delta_2$
of the two bulk fields through $2a = \Delta_2-\Delta_1$. The range of the variables
$x_i$ is the same as in eq.\ \eqref{eq:xrange} for $N=2$. If we set the parameter $a$
to zero, we recover the Casimir operator \eqref{eq:Caspq} with parameters
\eqref{eq:Caspqpar} for $q=0$ and $p \leq d-2$. Hence, the parameter $a$ may be
regarded as a deformation that exists for $q=0$.

If $p=d-1$, while $q=0$ as in the previous paragraph, we are dealing
with a correlator of two bulk fields in the presence of a boundary or
conformal interface. In this case $N = \min (d-p,q+2) = \min (1,2) = 1$ so that
there is a single cross-ratio only, as is well known from \cite{McAvity:1995zd}. The
Casimir operator takes the simple form
\begin{equation}
L^2 = H_{CS} + \epsilon_0 \,,\quad \epsilon_0 = \frac{d^2}{16}
\end{equation}
with parameters
\begin{align}
N&=1 \,,\quad k_1=1 - 2a -\frac{d}{2} \,,\quad k_2=\frac{d-1}{2} \, .
\end{align}
Note that the Calogero-Sutherland model from $N=1$ contains only two multiplicities.
The corresponding eigenvalue equation can be mapped to the hypergeometric differential
equation. Once again, for $a=0$ we recover the Casimir problem \eqref{eq:Caspq} for two
defects of dimension $p=d-1$ and $q=0$.
\medskip

For reference, we conclude this list of results with the case $p=q=0$ which is associated
with correlations of four scalar bulk fields and was studied within the context of
Calogero-Sutherland models in \cite{Isachenkov:2016gim,Schomerus:2016epl}. In this
case the Casimir operator is known to take the form
\begin{equation}
L^2 = \frac12 H'_{CS} + \epsilon_0 \,,\quad \epsilon_0 = \frac{d^2-2d+2}{8}
\end{equation}
with
\begin{align}\label{eq:Cas00par}
N&=2 \,,\quad k_1=-2b \,,\quad k_2=a+b+\frac12 \,,\quad k_3=\frac{d-2}{2} \, ,
\end{align}
where the parameters $2a= \Delta_2-\Delta_1$ and $2b=\Delta_3-\Delta_4$ are
determined by the conformal weights of four external scalar fields. We put a
prime $'$ on the Hamiltonian to indicate that it actually depends on two
variables $u_1$ and $u_2$ that are complex conjugates of each other and belong
to the range
\begin{equation} \label{eq:urange}
\Re u_i \in [0, \infty[ \, \quad \Im u_1 = - \Im u_2 \in [0,\pi[\ .
\end{equation}
In contrast to the previous cases, the gauge transformation is now given by
\begin{align}\label{eq:gauge2}
\omega'(u_1,u_2) &= \prod_{i=1}^{2} \sinh^{a+b-\frac12}\left(\frac{u_i}{2}\right)
\cosh^{-(a+b)-\frac12}\left(\frac{u_i}{2}\right) \sinh^{-\frac{d-2}{2}}
\left(\frac{u_1\pm u_2}{2}\right) \,,
\end{align}
and the eigenvalues $\epsilon'$ of the Calogero-Sutherland Hamiltonian $H'$ are
related to the conformal weight $\Delta$ and the spin $\ell$ of the intermediate
field by $\epsilon'=-\frac12 C_{\Delta,\ell} - 2\epsilon_0$.

Of course, when we send the two parameters $a$ and $b$ to $a=b=0$ we expect to
recover the Casimir problem \eqref{eq:Caspq} for $p=q=0$. This is indeed true but
it requires to perform a non-trivial linear transformation on the coordinates and
the multiplicities. We shall denote this transformation by $\sigma_2$. It maps the
coordinates $\tau_1$ and $\tau_2$ to $u_1$ and $u_2$ as
\begin{equation}\label{eq:ucoord}
\sigma_2: \quad u_1 = \frac{\tau_1+\tau_2}{2} \,,\quad u_2 = \frac{\tau_1-\tau_2}{2}
\end{equation}
and the multiplicities $k_1, k_2=0$ and $k_3$ to
\begin{equation}
\label{eq:volker}
\sigma_2: \quad k'_1 = 0 \,,\quad k'_2 = k_3 \,,\quad k'_3 = k_1 \,.
\end{equation}
We note that $\sigma_2$ maps the range \eqref{eq:xrange} of the variables $\tau_i$
to the range \eqref{eq:urange}. Let us stress that we defined the transformation $\sigma_2$
only on Calogero-Sutherland Hamiltonians \eqref{eq:HCS} with multiplicity $k_2=0$. It is
not difficult to verify that upon acting with $\sigma_2$ on the Hamiltonian \eqref{eq:HCS}
we obtain a Hamiltonian $H'_\textrm{CS}$ of the same form iff\footnote{Or,
	equivalently, $k_2-1 = 0$, but this is already captured by symmetry $\rho_2$ in table
	\ref{tab:sym}.} $k_2 = 0$ (up to an overall factor of 2) but with multiplicities $k'_i$ instead
of $k_i$. For the case of interest here, i.e.\ when $p=q=0$, the condition $k_2=0$ is
indeed satisfied as one can infer from eq.\ \eqref{eq:Caspqpar}. After applying the
transformation \eqref{eq:volker} to the multiplicities we find $(k'_1,k'_2,k'_3) =
 (0,1/2,d/2-1)$. As we have claimed, we end up with
the set of parameters \eqref{eq:Cas00par} for $a = b = 0$. This is what we wanted to
show.

As a small corollary of the previous discussion let us briefly mention that the
transformation  \eqref{eq:volker} can be inverted in case $N=2$ and $k_1 = 0$. On
the coordinates, the inverse reads
\begin{equation}\label{eq:ucoord2}
\sigma_1: \quad v_1 = \tau_1+\tau_2 \,,\quad v_2 = \tau_1-\tau_2 \,,
\end{equation}
while it acts on the multiplicities as
\begin{equation}\label{eq:pedro}
	\sigma_1: \quad k'_1 = k_3 \,,\quad k'_2 = 0 \,,\quad k'_3 = k_2 \,.
\end{equation}
The maps $\sigma_1$ and $\sigma_2$ describe two symmetries of Calogero-Sutherland model with
$k_1=0$ and $k_2=0$, respectively, that exist for $N=2$ only and act on multiplicities as well
as coordinates. These symmetries are not included in table \ref{tab:sym} but will play some role
in our discussion below. Unlike the dualities displayed in table \ref{tab:sym} which generalize
Euler-Pfaff symmetries of Gauss hypergeometric function, the transformations \eqref{eq:ucoord} and \eqref{eq:ucoord2} represent special cases of quadratic transformations of Calogero-Sutherland wave
functions, generalizing classical quadratic transformations of Gauss hypergeometric functions.\footnote{See also \cite{Koornwinder-quadratic} for further results and a state-of-art discussion
of quadratic transformations among wave functions in the trigonometric case and e.g.
\cite{Rains-Vazirani} for elliptically-deformed analogues.}

\subsection{Application: Relations between blocks}

Before we sketch how the results of the previous section are derived we want to pause for a
moment and discuss some immediate consequences that can be obtained from the equations alone
without detailed knowledge about their solutions.

\subsubsection{Relations between defect blocks with $q \neq 0$}

As we stressed before, the Calogero-Sutherland Hamiltonian \eqref{eq:HCS}, i.e. the quadratic Casimir operator for the block, possesses some obvious
symmetries which we listed in table \ref{tab:sym}.
In the previous subsection we have explained
how the coupling constants $k_i$ of the Calogero-Sutherland model are determined by the dimension
$p$ and $q$ of the two defects and the dimension $d$. Putting this together, we can rephrase the
symmetries from table \ref{tab:sym} in terms of the parameters $(p,q;d)$. The result is stated in
table \ref{tab:symdd}. The first two symmetry transformations $\varrho_1$ and $\varrho_2$ give rise
to non-trivial relations between the parameters while the third one acts trivially  on the
coupling constants of our Calogero-Sutherland model since $k_3 = 1/2 = 1-k_3 = k_3'$. Let us also
note that the reconstruction of $p,q$ and $d$  from the multiplicities is not unique since they
depend on $p$ and $q$ only through $N$ and $p-q$. The ambiguity is described by the following duality
\begin{equation}
\label{eq:abhijit}
p' = d-q-2 \,,\quad q' = d-p-2 \,,
\end{equation}
which we included as the final row of the table. It makes up for the trivial third row. As in table,
\ref{tab:sym}, the forth row describes a symmetry for which the action on parameters is accompanied
by a shift of coordinates $\tau_i \rightarrow \tau_i \pm i \pi$.
\begin{table}[h!]
	\centering
	\caption{The action of symmetries in table \ref{tab:sym} on the parameters $(p,q;d)$ that characterize
a configuration of two defects. As in table \ref{tab:sym} the symmetry transformation $\tilde \varrho$ is
accompanied by a shift of coordinates. The last row is new and results from the fact it is not possible to
reconstruct the parameters $(p,q;d)$ uniquely from the coupling constants $k_i$.}
	\begin{tabular}{l|c|c|c}
		 & $p'$ & $q'$ & $d'$ \\[2mm] \hline & & & \\[-3mm]
		$\varrho_1$ & $N+(p-q)-2$ & $N-2 $ & $4N-d+2(p-q)-2$ \label{eq:k1_defdef} \\[2mm]
		$\varrho_2$ & $N-(p-q)$ & $N-2$  & $4N-d$ \label{eq:k2_defdef} \\[2mm]
		$\varrho_3$ & $ p $ & $ q $ & $ d $ \label{eq:k3_defdef} \\[2mm]
		$ \tilde \varrho$ & $ 3N-d+(p-q)-2 $ & $ N-2 $ & $ 4N-d $ \label{eq:marco_defdef} \\[2mm]
		$\varrho_0$ & $d-q-2$ & $d-p-2$ & $d$ \label{eq:abhijit_defdef}
	\end{tabular}
\label{tab:symdd}
\end{table}
These innocent looking relations have remarkable consequences of which we have seen a very special
case before when we reviewed the results from \cite{Gadde:2016fbj}. Namely, in section 2.3 we
discussed the blocks for a two point function for defects of dimension $p=q = d-2$. If we plug
these values into the relation \eqref{eq:abhijit} we find $p'=0=q'$, i.e.\ the blocks for two
point functions of defects of dimension $p=d-2=q$ are related to four-point blocks of scalar
bulk fields. As we explained in the previous subsection, the relation between the two
Calogero-Sutherland problems involves the coordinate transformations \eqref{eq:ucoord} and
\begin{equation}\label{eq:CS4ptCoord}
	z = -\sinh^{-2}\left(\frac{u_1}{2}\right) \,,\quad \bar{z} = -\sinh^{-2}\left(\frac{u_2}{2}\right) \,.
\end{equation}
Using the relations \eqref{eq:xrange} and \eqref{eq:etatheta}, we recover the relation
\eqref{eq:rel3} observed in \cite{Gadde:2016fbj}. More generally, any relation between
Calogero-Sutherland models that can be obtained by applying one or several of the symmetries
in table \ref{tab:symdd} leads to a relation between solutions. In case one does not need to
apply the symmetry $\tilde \rho$, the Euclidean region of one system is mapped to the
Euclidean of the other and hence one can also match boundary conditions so that all
symmetries other than $\tilde \rho$ actually map blocks to blocks. Thereby, our table
\ref{tab:symdd} provides a vast generalization of eq.\ \eqref{eq:thirdrel}.

\subsubsection{Defect configurations with $q=0$ and four-point blocks}

The other two relations between defect blocks and those for scalar four-point functions that
we discussed in section 2.3 involve configurations with $q=0$. We have determined the coupling
constants of the associated Calogero-Sutherland model in eqs.\ \eqref{eq:Casp0par}. Once again
we can apply the symmetries from table \ref{tab:sym} to find the symmetry relations listed in
table \ref{tab:symdb}.
\begin{table}[h!]
	\centering
	\caption{The action of symmetries in table \ref{tab:sym} on the parameters $(p,a;d)$ that
characterize a configuration of two scalar bulk fields in the presence of a single defect. As
in table \ref{tab:sym} the symmetry transformation $\tilde \varrho$ is accompanied by a shift
of coordinates.}
	\begin{tabular}{l|c|c|c}
		 & $p'$ & $a'$ & $d'$ \\[2mm] \hline & & & \\[-3mm]
		$\varrho_1$ & $p$ & $a$ & $2p-d+6$ \label{eq:k1_blkdef} \\[2mm]
		$ \varrho_2$ & $2-p$ & $a$  & $8-d$ \label{eq:k2_blkdef} \\[2mm]
		$\varrho_3$ & $ p $ & $-a$ & $ d $ \label{eq:k3_blkdef} \\[2mm]
		$ \tilde \varrho$ & $ 4-d+p $ & $ a $ & $ 8-d $ \label{eq:marco_blkdef}
	\end{tabular}
\label{tab:symdb}
\end{table}

Let us re-derive and generalize the relation \eqref{eq:secondrel} between two identical scalars
in the presence of a line defect in $d=4$ dimensions and scalar four-point blocks from
\cite{Liendo:2016ymz}. We actually want to consider two scalar fields whose weights differ by
$a=(\Delta_2-\Delta_1)/2$ in the presence of a ($d/2-1$)-dimensional defect in $d$ dimensions.
According to the general results, the corresponding Calogero-Sutherland model has $N=2$
coordinates $\tau_1,\tau_2$ and its coupling constants are determined by the parameters
$(p,a;d)=(\frac{d}{2}-1,a;d)$ of the configuration through eq.\ \eqref{eq:Casp0par}, i.e.\
$k'_1=0$. This means that we can apply the symmetry $\sigma_1$ that we introduced at the
end of the previous subsection. The resulting triple of multiplicities can be interpreted as a
set of multiplicities \eqref{eq:Cas00par} in the Calogero-Sutherland model for scalar four-point
block with weights
$$ a'=\frac12 (\Delta'_2-\Delta'_1) = -\frac14+\frac{a}{2} \quad , \quad
b'=\frac12 (\Delta'_3-\Delta'_4) =-\frac14-\frac{a}{2}$$
in a $(d/2+1)$-dimensional Euclidean space. In order to compare with the duality \eqref{eq:secondrel}
found in \cite{Liendo:2016ymz} we need to flip the sign of $a'$ by applying $\tilde{\varrho}$. So, in
order to match the parameters we have applied the symmetry transformations $\sigma_1$ and
$\tilde \varrho$.

Let us now see how these transformations act on the coordinates. Since both $\sigma_1$ and $\tilde
\varrho$ act on them non-trivially, the map between the parameters $x, \bar{x}$ of the original
configuration and the cross-ratios $\gamma, \bar{\gamma}$ of the four-point blocks will be non-trivial as well.
Recall the relations \eqref{eq:xxbtheta} and \eqref{eq:xrange} between the coordinates $x, \bar{x}$
and our coordinates $\tau_1$, $\tau_2$. After applying $\sigma_1$ we pass to the cross-ratios $y,
\bar{y}$ using eq.\ \eqref{eq:CS4ptCoord} to obtain
\begin{equation}
	y = -\frac{(1-x)^2}{4x} \,,\quad \bar{y} = -\frac{(1-\bar{x})^2}{4\bar{x}} \,.
\end{equation}
Next we need to apply $\tilde{\varrho}$, i.e.\ shift the coordinates $v_1,v_2$ by $i \pi$ to obtain\footnote{We need to exploit the $2\pi i$-periodicity of the potential and shift $v_2$ by $-2\pi i$ in order to ensure that $v_1$, $v_2$ stay complex conjugates.}
\begin{equation}
	\gamma = \frac{y}{y-1} = \left(\frac{1-x}{1+x}\right)^2 \,,\quad
\bar{\gamma} = \frac{\bar{y}}{\bar{y}-1} = \left(\frac{1-\bar{x}}{1+\bar{x}}\right)^2 \,,
\end{equation}
which is precisely the relation between the relevant cross-ratios that was found in \cite{Liendo:2016ymz}.

It remains to identify the weight and spin of the exchanged field in the scalar four-point blocks. In order
to do so we only need to impose the correct asymptotics of the blocks on both sides. This is done in two
steps. First, we obtain the gauge transformation between the defect block $f$ and the corresponding
four-point block $g$ by using \eqref{eq:gauge1} and \eqref{eq:gauge2}. Then we impose the limit \eqref{eq:defblocknorm}\footnote{Note that the normalization differs from \cite{Billo:2016cpy}, i.e.\ $f^{there}=2^{-\ell}f^{here}$. For the scalar four-point blocks, we adopt a normalization of \cite{Caron-Huot:2017vep}. To switch to conventions of \cite{Dolan:2011dv, Isachenkov:2017qgn}, one should multiply our scalar blocks by $(d/2-1)_{\ell'}/(d-2)_{\ell'}$.}
\begin{align}
\cblock{f}{p,a,d}{\Delta,\ell}{x,\bar{x}} &\stackrel{x,\bar{x}\rightarrow 1}{\longrightarrow}
[(1-x)(1-\bar{x})]^{\frac{\Delta-\ell}{2}} (2-x-\bar{x})^\ell \label{eq:asymp_defect} \,, \\[2mm]
\cblock{g}{a',b',d'}{\Delta',\ell'}{z,\bar{z}} &\stackrel{z,\bar{z}\rightarrow 0}{\longrightarrow}
(z\bar{z})^{\frac{\Delta'-\ell'}{2}} (z+\bar{z})^{\ell'} \label{eq:asymp_4pt} \,,
\end{align}
which fixes $\Delta', \ell'$. The final result that we obtain from our symmetries and the comparison of
asymptotics is
\begin{align}  \label{eq:genrel1}
\cblock{f}{\frac{d}{2}-1,a,d}{\Delta,\ell}{x,\bar{x}} &= (-1)^{-\frac{\ell}{2}} 2^{\Delta} (y\bar{y})^{-\frac14}
\cblock{g}{-\frac14+\frac{a}{2},-\frac14-\frac{a}{2},\frac{d}{2}+1}{\frac{\Delta+1}{2}, \frac{\ell}{2}}{y,\bar{y}} \\[2mm]
&= 2^{\Delta} (\gamma\bar{\gamma})^{-\frac14} \left[(1-\gamma)(1-\bar\gamma)\right]^{-\frac{a}{2}}
 \cblock{g}{\frac14-\frac{a}{2},-\frac14-\frac{a}{2},\frac{d}{2}+1}{\frac{\Delta+1}{2}, \frac{\ell}{2}}{\gamma,\bar{\gamma}}
 \, .
\end{align}
The first line corresponds to the application of $\sigma_1$ only. To pass to the second line we used
that the scalar four-point blocks transform under $\tilde{\varrho}$ as
\begin{equation}
	\cblock{g}{a',b',d'}{\Delta',\ell'}{z,\bar{z}} = (-1)^{\ell'} \left[(1-z)(1-\bar{z})\right]^{-b'} \cblock{g}{-a',b',d'}{\Delta',\ell'}{\frac{z}{z-1},\frac{\bar{z}}{\bar{z}-1}}
\end{equation}
for integer $\ell'$. The resulting formula indeed reduces to eq.\ \eqref{eq:secondrel} when we choose
$d=4$ and $a=0$ and hence provides a rather non-trivial extension. There are three other dualities
between defect and four-point blocks that can be derived along the same route, one more involving
the symmetry $\sigma_1$,
\begin{align}  \label{eq:genrel2}
\cblock{f}{p,a,d=4}{\Delta,\ell}{x,\bar{x}} &= (-1)^{-\frac{\ell-p+1}{2}} 2^{\Delta} (y\bar{y})^{-\frac14}
\left| \sqrt\frac{y-1}{y} - \sqrt\frac{\bar{y}-1}{\bar{y}} \right|^{p-1} \notag
\\[2mm]  &\qquad\times
\cblock{g}{-\frac14+\frac{a}{2},-\frac14-\frac{a}{2},p+2}{\frac{\Delta+p}{2}, \frac{\ell-p+1}{2}}{y,\bar{y}}
\, ,
\end{align}
and two involving $\sigma_2$,
\begin{align} \label{eq:genrel3}
\cblock{f}{p=0,a,d}{\Delta,\ell}{x,\bar{x}} &= (x\bar{x})^{\frac{a}{2}} \cblock{g}{a,0,d}{\Delta,\ell}{1-x,1-\bar{x}} \,, \\[2mm]
\cblock{f}{p=2,a,d}{\Delta,\ell}{x,\bar{x}} &= \frac{(1-x)(1-\bar{x})}{x\bar{x}-1} (x\bar{x})^{\frac{a}{2}}
\cblock{g}{a,0,d-2}{\Delta-1,\ell+1}{1-x,1-\bar{x}} \,.
\label{eq:genrel4}
\end{align}
Note that eq.\ \eqref{eq:genrel3} applies to $p=0$ and hence it maps four-point blocks to four-point blocks,
as was already discussed for $a=0$ in the previous subsection. The prefactor $(x\bar{x})^{\frac{a}{2}}$ on
the right hand side stems from different gauge choices used in the literature.

Finally, let us comment on the duality \eqref{eq:firstrel} from \cite{Billo:2016cpy} that relates two-point
functions in presence of a $d-2$-dimensional defect to four-point blocks in the same dimension. It is not
difficult to identify the symmetries that are needed to relate the parameters on the left and the right
hand side. In fact, one simply needs to apply the symmetry $\tilde{\varrho}$ in table \ref{tab:symdb}
before passing to the four-point case using $\sigma_2$. Allowing once again for non-vanishing $a$ one
obtains
\begin{equation} \label{eq:almostrel}
	f(p=d-2,a,d) \sim g(a,0,d) \quad \text{and} \quad f(p=d-4,a,d) \sim g(a,0,d+2) \,.
\end{equation}
Here, we have only displayed the parameters in the first row of the defect blocks $f$ and the four-point
blocks $g$, i.e.\ we suppressed the dependence on conformal weights and cross-ratios. As in our discussion
above, one can apply the symmetries to the cross ratios only to find that the resulting transformation does
not map the Euclidean domain of the defect cross-ratios to the Euclidean domain of the four-point block, but instead to a Lorentzian domain. Hence,
eq.\ \eqref{eq:almostrel} does not provide a relation between blocks but involves analytic continuation (see section 2.3). Nevertheless, we will be able to
construct the relevant defect blocks directly in section 5, without passing through four-point blocks.
Let us stress again that in this subsection we did not only recover all previously known relations between
blocks form the symmetries of the Calogero-Sutherland model, but we also extended them vastly, see in
particular the relations \eqref{eq:genrel1}-\eqref{eq:genrel4}.

\subsection{Derivation of results}

In the final subsection we want to sketch the derivation of the results we presented and
discussed in the subsection 3.1. Many more details can be found in \cite{Schomerus:2016epl}
where similar results were derived for the blocks of four scalar bulk fields. Here we
shall briefly introduce some relevant background from group theory before we define
the space of conformal blocks and evaluate the conformal Casimir on this space. The
subsection concludes with a discussion of the coordinates.

As we have stated before, a $p$-dimensional conformal defect breaks the conformal group
$G = \SO{1,d+1}$ down to the subgroup
\begin{equation}
G_p = \SO{1,p+1} \times \SO{d-p} \ \subset \ G \ .
\end{equation}
Here, the first factor describes conformal transformations of the world-volume of
the defect and the second factor accounts for rotations of the transverse space.
Elements of the $d$-dimensional conformal group $G$ that are not contained in the
subgroup $G_p$ act as transformations on the defect. The number of such non-trivial
transformations is given by the dimension of the quotient $G / G_p$,
\begin{equation}
\dim G/G_p = (p+2)(d-p) \ .
\end{equation}
For $p=0$, the defect $D_{p=0}$ consists of a pair of points and the $2d$-dimensional
quotient $G/G_0$ describes their configuration space. When we set $p=d-1$, i.e.\ consider
a defect of codimension $d-p=1$, the quotient $G/G_p$ has dimension $\dim G/G_{d-1} =
d+1$. A $(d-1)$-dimensional conformal defect is localized along a sphere in the
$d$-dimensional background and the $d+1$ parameters provided by the surface
$G/G_{d-1}$ represent the position of its centre and the radius.

In order to define the space of blocks we must first choose two finite dimensional
irreducible (unitary) representations $\pi_L$ and $\pi_R$ of the groups $G_p$ and
$G_q$. Here we shall restrict to scalar blocks from the very beginning which means
that $\pi_L$ and $\pi_R$ are assumed to be one-dimensional. For $p,q \neq 0$,
the only one-dimensional representation is the trivial one. Only if either $q$ or
even $p$ and $q$ vanish, one can have a non-trivial one-dimensional representation
for which the generator of dilations is represented by a complex number. We shall
denote these parameters by $b$ and $a$, respectively. If $p,q \neq 0$ the space of
conformal blocks is given by
\begin{equation}
 \Gamma_{pq}  \ = \  \{ \, f:G \rightarrow \mathbb{C} \, | \,
f(h_L g h_R) =  f(g) \, ; \, h_L \in G_p , h_R \in G_q\, \}\ ,
\end{equation}
i.e.\ it consists of all complex valued functions on the conformal group that are
invariant with respect to left translations by elements $h_L \in G_p$ and to right
right translations by elements $h_R \in G_q$. When $q=0$ but $p\neq 0$, translations
with elements
\begin{equation}\label{SO(1,1)par}
d(\lambda) = \left( \begin{array}{cc} \cosh\lambda & \sinh\lambda \\
\sinh\lambda & \cosh\lambda \end{array} \right)\
\end{equation}
of the subgroup $D = \SO{1,1} \in G_0$ are accompanied by a non-trivial phase shift
 \begin{equation}
 \Gamma^a_{p}  \ = \  \{ \, f:G \rightarrow \mathbb{C} \, | \,
f(h_L g d h'_R) =  e^{-2a\lambda} f(g) \, ; \, h_L \in G_p , h'_R \in \SO{d}\, \}\ .
\end{equation}
In case both $p$ and $q$ vanish, finally, the resulting space of scalar four-point
blocks is given by \cite{Schomerus:2016epl}
\begin{equation}
 \Gamma^{ba}  \ = \  \{ \, f:G \rightarrow \mathbb{C} \, | \,
f(h_L g d h'_R) =  e^{2(b-a)\lambda} f(g) \, ; \, h'_L , h'_R \in \SO{d}\, \}\ .
\end{equation}
In all three cases, the elements of the space $\Gamma$ are uniquely determined by
the values they take on the double quotient $G_p\backslash G/ G_q$. This two-sided
coset parametrizes the space of cross-ratios. The precise relation between cross-ratios and coordinates on the conformal groups will be discussed below. For the
moment let us only check that the double quotient is $N$-dimensional. In order to
see that, we anticipate from our discussion of coordinates below that a point on
the double quotient is stabilized by the subgroup
\begin{equation}
 B_{pq} = \SO{p-q} \times \SO{|d-p-q-2|} \subset G_p, G_q \subset G\ .
\end{equation}
Once this is taken into account, it is is straightforward to compute the
dimension of the double coset space,
$$ \dim G_p \backslash G / G_q = \dim G - \dim G_p - \dim G_q + \dim B_{pq}
= N \ .
$$
All this is valid for any choice of $p,q$ including $p=q=0$. In the latter case, the
double coset coincides with the one that was introduced in the context of scalar
four-point blocks \cite{Schomerus:2016epl}.

\begin{figure}[htb]
\centering
\includegraphics[scale=.3]{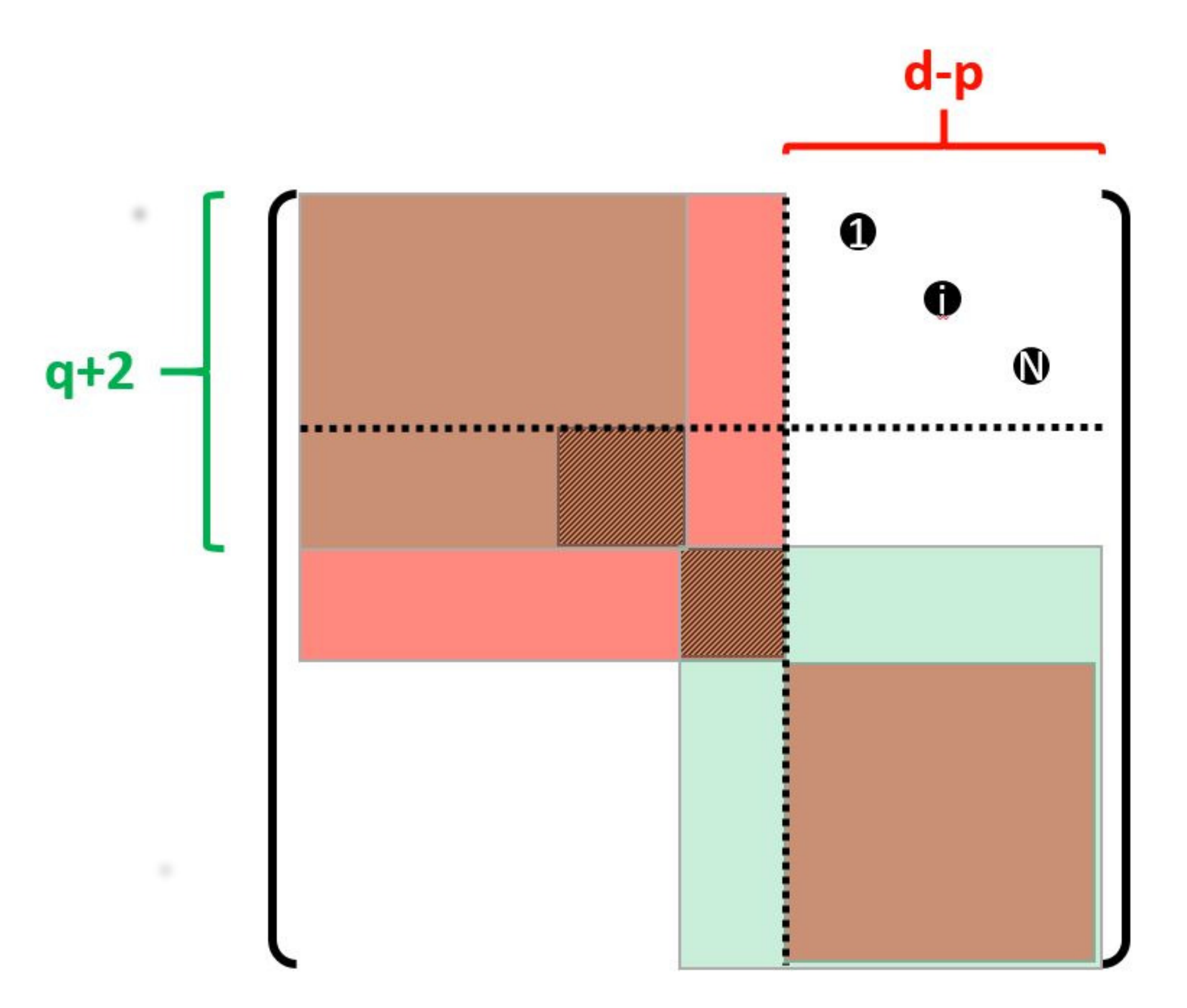}
\caption{The figure illustrates our choice of coordinates on the conformal
group. The blocks in red/green correspond to the left/right group $G_p/G_q$
while the additional generators $M_{i-1,p+1+i}$ are represented by block
dots. The subgroup $B_{pq}$ of elements that commute with $M_{i-1,p+1+i},
i=1, \dots, N$ is shown as the shaded area. Obviously, it is contained in
the intersection of $G_p$ and $G_q$ (brown area).}
\label{fig:coordinates}
\end{figure}

The space $\Gamma$ of conformal blocks comes equipped with an action of several differential
operators. In fact, the Casimir elements of the conformal group $G$ give rise to differential
operators for functions on the conformal group with the usual Laplacian associated to the
quadratic Casimir element. Higher order differential operators come with the higher order
Casimir elements. These differential operators on the group commute with both left and
right  translation and hence they descend to a set of commuting differential operators
on the space $\Gamma$. By definition conformal blocks are eigenfunctions of these
differential operators. In deriving the results of the previous subsection our main task
is to evaluate the quadratic Casimir element on the quotient $G_p\backslash G/G_q$. This
is facilitated by a choice of coordinates on  the conformal group that is adapted to the
geometrical setup. More precisely, we shall parametrize elements $g \in G$ of the
conformal group as
\begin{equation}\label{eq:coordinates}
g = h'_L a(\tau) h_R\  \quad  h_R \in G_q \, , \ h'_L \in G_p/B_{pq}\ .
\end{equation}
The choice of coordinates for elements $h_R \in G_q$ of the subgroup $G_q$ is not
important. In order to parametrize the subgroup $G_p$ one should first choose
coordinates on the subgroup $B_{pq}$ and then extend these to coordinates of
$G_p$. Elements $h'_L$ of the $(\dim G_p - \dim B_{pq})$-dimensional quotient
$G_p/B_{pq}$ do not depend on the coordinates on $B_{pq}$. In order to factorise
elements $g$ of the conformal group as in eq.\ \eqref{eq:coordinates}, we need
$N$ additional coordinates which parametrize the factor $a= a(\tau)$ in the
middle. This takes the form
$$ a(\tau) = e^{\tau_i M_{i-1,p+1+i}} \in A_{pq} \quad \textrm{ where} \ M_{i-1,p+1+i}
\, , \quad i = 1, \dots, N $$
are the usual generators of $\SO{1,d+1}$. In particular, the generators $M_{i-1,p+1+i}$
with $i \geq 3$ are generators of rotations in the $(i-2,p+i)$-plane while
$$ M_{0,p+2} = \frac12 \left(P_{p+2} - K_{p+2}\right) \quad , \quad
M_{1,p+3} = \frac12\left(P_{p+3} + K_{p+3}\right)  $$
are linear combinations of infinitesimal translations and special conformal
transformations. The various subgroups and the generators $M_{i-1,p+1+i}$ of
the torus $A$ are illustrated in figure \ref{fig:coordinates}. Let us note
that the generators $M_{i-1,p+1-i}$ commute with elements in the subgroup
$B_{pq}$, a result we anticipated above.

Once we have fixed our coordinates on $G$ it is straightforward to compute first
the metric and then the Laplace-Beltrami operator $\Delta_\textrm{LB}$ on $G$. The
resulting expression is a second order differential operator that contains derivatives
with respect to all the coordinates on the conformal group, including the coordinates
$\tau_i$ on the torus $A_{pq}$ and the parameters $\lambda_R$ and $\lambda_L$ on the
subgroups $D = \SO{1,1}$ of dilations in case $q=0$ or $p=q=0$. In order to descend to
the space of conformal blocks we have to set all other derivatives to zero so that
we end up with a second order differential operator $\Delta^A_\textrm{LB}$ in $\tau_i$.
In case $q=0$ or $p=q=0$ the derivatives with respect to $\lambda_R$ and $\lambda_L$ are
replaced by $-2a$ and $2b$, respectively. The operator $\Delta^A_\textrm{LB}$ still
turns out to contain some first order terms. The latter can be removed by an appropriate
``gauge transformation'' \eqref{eq:gauge1}.
The Casimir operators $L^2$ we listed in the previous subsection are given by
\begin{equation}
L^2 =  \omega^{-1} \, \Delta^A_\textrm{LB}\,  \omega\ .
\end{equation}
It remains to relate the group theoretic variables $\tau_i$ we introduced through
our parametrization of the conformal group $G$ to the cross-ratios. As we explained
above, the location of the defect operators $\mathcal{D}^{(p)}(P_\alpha)$ and
$\mathcal{D}^{(q)}(Q_\beta)$ can be characterized by a set of orthonormal vectors
$P_\alpha, \alpha = p+2, \dots ,d+1,$ and $Q_\beta, \beta = q+2, \dots, d+1,$ which
are transverse to the defect in embedding space, respectively. We can complete
these two sets to an orthonormal basis $\mathcal{P}$, $\mathcal{Q}$ of the full
embedding space by adding vectors $\tilde P_\alpha, \alpha =0, 1,\dots,p+1,$ and
$\tilde Q_\beta, \beta = 0,1, \dots,q+1$. Let us now combine these systems of
orthonormal vectors into two matrices
\begin{equation}
\mathcal{P}=(\tilde{P},P)\in G=SO(1,d+1) \,,\quad \mathcal{Q}=(\tilde{Q},Q)\in G \,.
\end{equation}
By construction, both $\mathcal{P}$ and $\mathcal{Q}$ carry a left action of the
conformal group (since the columns are vectors in embedding space) and a right
with respect to $G_p$ and $G_q$, respectively. The latter respects the split of
the columns into vectors tangential and transverse to the defect. For the two
$\SO{1,d+1}$ matrices $\mathcal{P}$ and $\mathcal{Q}$ we can now form the matrix
$\mathcal{P}^T\mathcal{Q} \in \SO{1,d+1}$. Obviously, $\mathcal{P}^T \mathcal{Q}$
is invariant under conformal transformations, but it transforms non-trivially under
the action of $G_p$ and $G_q$. In this way, any configuration of two defects of
dimension $p$ and $q$ gives rise to an orbit $G_p \mathcal{P}^T \mathcal{Q} G_q$
in the double quotient $G_p\backslash G / G_q$.

In section 2 we considered the matrix $M = P^TQ$ in order to construct the cross-ratios
$\eta_i$ of the defect configurations. Now we see that $M$ appears as the lower
right matrix block of the matrix $a(\tau)$ we introduced in eq.\
\eqref{eq:coordinates}. From the explicit construction in terms of the generators
$M_{i-1,p+1+i}$ we can see that the lower right corner of $a(\tau)$ takes the
form
\begin{equation}
	\left(\begin{array}{@{}ccccc|c@{}}
		\cosh\frac{\tau_1}{2} & & & & & \\
		& \cosh\frac{\tau_2}{2} & & & & \\
		& & \ddots & & & 0 \\
		& & & \cosh\frac{\tau_N}{2} & & \\
		& & & & I &
	\end{array}\right) \,.
\end{equation}
Comparison with our discussion of the cross-ratios allows us to read off the relation
\eqref{eq:xrange} between the group theoretic variables and cross-ratios.

The last task is to relate the Calogero-Sutherland  eigenfunctions to the conformal blocks.
In case of $p,q>0$, the Casimir equation for the correlator is the same as for the block
(see eq.\ \eqref{eq:CPWexp}). Hence we just need to undo the gauge transformation \eqref{eq:gauge1}
and arrive at eq.\ \eqref{eq:cb_wavefct}. In case the defect configuration includes local fields,
i.e.\  when $q=0$ or $p,q=0$, the Casimir equations have been worked out \cite{Dolan:2003hv,Billo:2016cpy}
and we arrive at eqs.\ \eqref{eq:gauge1} and \eqref{eq:gauge2}, respectively. This concludes the brief
sketch of the derivation of the results we listed in the first subsection. The interested reader can
find many more details in \cite{Schomerus:2016epl} where the case of scalar four-point blocks is
analysed.

\section{Calogero-Sutherland scattering states}

Here we present a review of the solution theory. We introduce the fundamental domain of
the Calogero-Sutherland problem and its fundamental (monodromy) group,  Harish-Chandra
scattering states, the monodromy representations and physical (monodromy free) wave
functions.

\subsection{Symmetries and fundamental domain}

It is useful to consider the Calogero-Sutherland potential \eqref{eq:HCS} as a function of
$N$ complex variables first and to impose reality conditions a bit later. As a function
of complex coordinates $\tau_i \in \mathbb{C}$, the potential possesses a few important
symmetries. These include independent shifts of the coordinates $\tau_i$ by $2\pi i$ in
the imaginary direction as well as two types of reflections, namely the inversion
symmetries $\tau_i \leftrightarrow - \tau_i$ and the particle exchange symmetry $\tau_i
\leftrightarrow \tau_j$. Together these form a non-abelian group that mathematicians
refer to as affine Weyl group $\mathcal{W}_N$. The reflections actually generate a
usual Weyl group and the shifts make this affine. The affine Weyl group is known to
possess a so-called Coxeter representation through $N+1$ generators $w_i, i=0, \dots,
N$ with relations
\begin{eqnarray}
\  w_i w_j & = & w_j w_i \quad
\textrm{for} \ |i-j|\geq 2  \, ,  \label{eq:w1} \\[2mm]
w_i w_{i+1} w_i  & = & w_{i+1} w_i w_{i+1} \quad \textrm{for} \ i=1, \dots\, N-2 \, ,
\label{eq:w2} \\[2mm]
w_{0} w_{1} w_{0} w_{1} = w_{1} w_{0} w_{1} w_{0}
& , & w_{N-1} w_{N} w_{N-1} w_{N} = w_{N} w_{N-1} w_{N} w_{N-1} \  .
\label{eq:w3}
\end{eqnarray}
and
\begin{equation}\label{eq:w4}
w_i^2 = 1 \quad \textit{ for all} \quad i=0, \dots, N \ .
\end{equation}
In this presentation of the affine Weyl group, the generators of the shifts in the
imaginary direction are a bit hidden, but they can be reconstructed from the $w_i$,
see \cite{van1983homotopy,HeckmanBook}.

\begin{figure}[htb]
	\centering
	\includegraphics[scale=0.5]{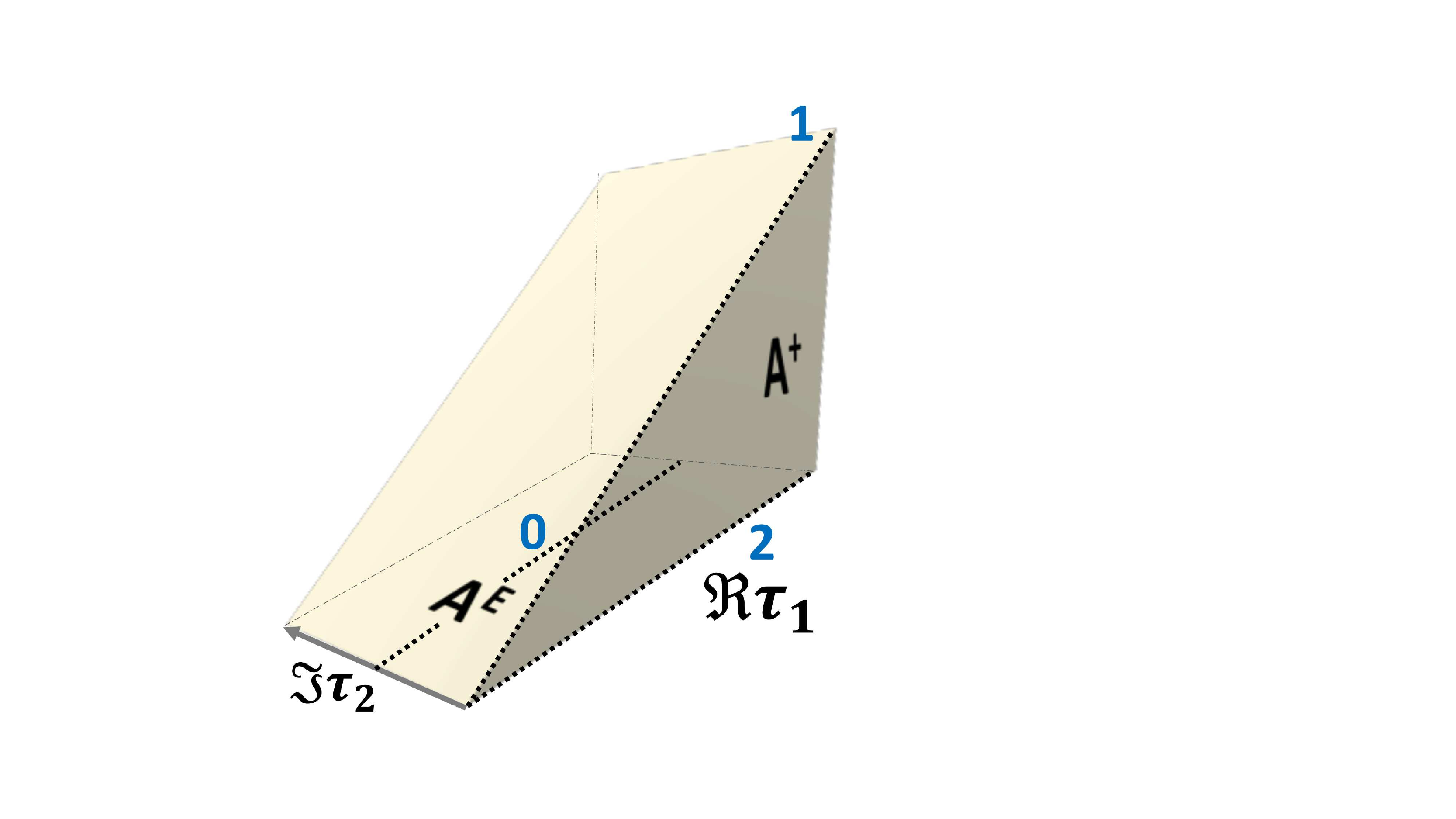}
	\caption{A 3-dimensional slice of the fundamental domain $D_2$ for the
		\textit{BC}$_2$ Calogero-Sutherland model in $\tau$-space with $\Im \tau_1=0$.
        Front and back side of the wedge should be identified. The fixed points
        (walls) under the action of $w_2$ and $w_1$ are shown as bold dashed lines.
        Fixed points of $w_2$ fall into two disconnected components which carry the
        labels $0$ and $2$. The shaded area in front is the Weyl chamber $A_2^+$. It
        is bounded by the walls $\wall_1$ and $\wall_2$. The subset $A^E_2$ is the
        2-dimensional semi-infinite strip of width $2\pi$ on the bottom of the wedge.
        It is bounded by the wall $\wall_2$, whereas wall $\wall_0$ cuts through its middle.}
	\label{fig:D2}
\end{figure}

The fundamental domain for the Calogero-Sutherland model is given by the quotient of the
configuration space $\mathbb{C}^N$ with respect to the symmetries, i.e.
\begin{equation}
D_N = \mathbb{C}^N / \mathcal{W}_N \ .
\end{equation}
We have depicted a 3-dimensional projection of the fundamental domain for
$N=2$ in figure \ref{fig:D2}. Inside the wedge-shaped domain, the Calogero-Sutherland
potential is finite but it diverges along the edges. We will refer to the hyperplanes
of singularities as ``walls'' of the Calogero-Sutherland model. It turns out that the
model possesses $N+1$ different walls $\omega_i, i=0, \dots, N$, one for each generator
$w_i$ of the affine Weyl group. For $N=2$ there are three such walls which are shown
in figure \ref{fig:D2}. The possible real domains $A_N^\alpha$ of the model are given
by the various faces of the domain $D_N$. Mathematicians usually  study the Schroedinger
problem in the real wedge $A_N^+$ which is given by $\tau_i \in \mathbb{R}$ with
$\tau_i > \tau_j > 0$ for all $i < j$.
\medskip

The fundamental group $\pi_1(D_N)$ of the fundamental domain plays an important
role in Calogero-Sutherland theory. It is generated by $N+1$ generators $g_i$ subject
to the relations \eqref{eq:w1}-\eqref{eq:w3} with $w_i$ replaced by $g_i$. On the
other hand, the generators $g_i$ of the fundamental group do not satisfy relation
\eqref{eq:w4}. The fundamental group of the domain $D_N$ is also referred to as
affine braid group. Its relation to the affine Weyl group is like the relation
between the braid group and the permutation group. Let us note that the
generators $w_a, a=1, \dots, N-1$ generate a subgroup $S_N \subset W_N$ of the
affine Weyl group that is isomorphic to the symmetric group $S_N$. The corresponding
generators $g_a, a=1, \dots, N-1,$ within the monodromy group generate Artin's
braid group. In addition, the full monodromy contains two more generators, $g_0$
and $g_N$ which satisfy some fourth order `reflection type' equations with $g_1$
and $g_{N-1}$, respectively.

\subsection{Harish-Chandra scattering states}

Before we enter our discussion of wave functions, it is advantageous to introduce
a bit of notation. We shall denote by $e_i, i=1, \dots, N,$ the $i^\textrm{th}$
unit vector in $\mathbb{C}^N$, i.e. the vector that is zero everywhere except in
the $i^\textrm{th}$ entry which is one instead. From these unit vectors we build
the following set $\Sigma^+$ of vectors in $\mathbb{C}^N$,
\begin{equation}
\Sigma^+ = \{e_i, 2 e_i, e_i\pm e_j|1 \leq i,j
\leq N; i < j \}\ .
\end{equation}
As one can easily count, the set contains $N(N+1)$ elements. Looking back at our
Calogero-Sutherland potentials we observe that they contain one summand for each
element in $\Sigma^+$. In fact, we can also write the potential as
\begin{equation}\label{CSgen}
V^{\text{CS}}(\tau_i) = \sum_{\alpha\in \Sigma^+}
\frac{k_\alpha(k_\alpha+ 2k_{2\alpha}-1)\langle \alpha,\alpha\rangle}
{4\sinh^2 \frac{\langle \alpha,\tau \rangle}{2} }\ .
\end{equation}
where the scalar product $\langle \cdot , \cdot \rangle$ is normalized such that
$\langle e_i , e_j \rangle = \delta_{i,j}$ and we assembled all the coordinates
$\tau_i \in \mathbb{C}$ into a vector $\tau = \sum_i \tau_i e_i$   with
$$ k_{e_i} = k_1 \quad , \quad k_{2e_i} = k_2 \quad , \quad k_{e_i\pm e_j} = k_3 \ .
$$
Let us agree to extend the definition of $k_\alpha$ to arbitrary elements $\alpha
\in \mathbb{R}^N$ such that is vanishes whenever $\alpha \not \in \Sigma^+$. Just
as in the case of the potential, many formulas below will turn out to become much
simpler when written as sums or products over the set $\Sigma^+$.

With these notations set up let us come to our main subject, namely the study of
wave functions. Since the Calogero-Sutherland potential falls off at $\tau_i \rightarrow
\infty$, any wave function becomes a superposition of plane waves in this asymptotic
regime. In mathematics is it customary to factor off the ground state wave function
of the trigonometric Calogero-Sutherland model, i.e.\ of the Hamiltonian that is
obtained when all the $\tau_i$ are purely imaginary. This ground state wave function
$\Theta$ is explicitly known,
\begin{equation} \label{eq:Thetadef}
\Theta(\tau_i) = \ \prod_{\alpha \in \Sigma^+}
\left( 2 \sinh \frac{\langle\alpha,\tau\rangle}{2}\right)^{k_\alpha}\ .
\end{equation}
For the wave function of the the Calogero-Sutherland model on the domain $A_N^+$ we
make the Ansatz
\begin{equation}
\Psi(\lambda,k;\tau) = \Theta(k;\tau) \Phi(\lambda,k;\tau)\ .
\end{equation}
Let us note in passing that the function $\Theta(k,\tau)$ possesses the following
asymptotics for large $\tau$,
\begin{eqnarray} \label{eq:Thetaasym}
\Theta(k;\tau) & \sim &   e^{\langle \rho_k,\tau\rangle } + \dots \quad \textit{where} \quad\\[2mm]
\rho_k &  := &  \left(\frac{k_1}{2} + k_2+(N-1)k_3, \frac{k_1}{2} + k_2+(N-2)k_3, \dots,
\frac{k_1}{2} + k_2 \right)\ . \nonumber
\end{eqnarray}
So-called Harish-Chandra wave functions $\Phi(\lambdan,k;\tau)$ are $W_N$ symmetric solutions
of the Calogero-Sutherland Hamiltonian for which $\Phi$ possesses the following simple
asymptotic behavior
\begin{equation}\label{HCasym}
\Phi(\lambdan,k;\tau)\  \sim \ e^{\langle \lambda - \rho_k, \tau\rangle}
+ \dots \ \mbox{ for } \ \tau \rightarrow \infty\ \mathrm{in} \ A^+_N = \textit{WC}_N
\end{equation}
where $\lambda = \sum_i \lambda_i e_i$ and $\tau \rightarrow \infty$ in $A^+_N$ means that
all components become large while preserving the order $\tau_N < \tau_{N-1} < \dots < \tau_1$.
Imposing $W_N$ symmetry implies that as a function of
$\tau_i$, $\Phi$ is reflection symmetric and invariant under any permutation of the $\tau_i$.
The  condition \eqref{HCasym} selects a unique solution of the scattering problem
describing a single plane wave. It is analytic in the wedge $A^+_N$. The corresponding
eigenvalue of the Calogero-Sutherland Hamiltonian is
given by
$$ \varepsilon = \varepsilon(\lambda) = - \sum \lambda_i^2 \ . $$
When we required the Harish-Chandra functions to be symmetric, we used the action
of the Weyl group $W_N$ on the coordinate space. On the other hand, the Weyl group
also acts in a natural way on the asymptotic data $\lambda$ of the Harish-Chandra
functions by sending any choice of $\lambda$ through a sequence of Weyl reflections
to $w \lambda, w \in W_N$. In particular, the generators $w_j, j=1, \dots, N$ act as
\begin{equation}\label{momenta-reflections}
w_a \lambda_i = \delta_{a+1,i}\lambda_{i-1}+\left(1-\delta_{a,i}\right)\left(1-\delta_{a+1,i}\right)
\lambda_i+ \delta_{a,i}\lambda_{i+1} \quad , \quad w_N \lambda_i = (-1)^{\delta_{N,i}}
\lambda_i\
\end{equation}
for $a = 1, \dots, N-1$ and $i = 1, \dots, N$. Since the eigenvalue $\varepsilon$
is invariant under exchange and reflection of the momenta $\lambda_i$, our
Harish-Chandra functions come in families. For generic choices of $\lambda$, one
obtains $|W_N| = N! 2^N$ solutions $\Phi(w\lambda,k;\tau)$ which all possess the same
eigenvalue of the Hamiltonian.

At least for sufficiently generic values of the momenta,\footnote{A precise formulation
of the condition is stated in \cite{HeckmanBook}.} Harish-Chandra functions possess a series
expansion in the variables $x_i = \exp \tau_i$
\begin{align}\label{HCh-Heckman-def}
\Phi(\lambdan,k;\tau) = \sum_{\mu\in Q_+} \Gamma_{\mu}
(\lambda,k)e^{\langle \lambda-\rho_k-\mu,\tau\rangle}, \quad \quad  \Gamma_{0}(\lambda,k)=1,
\end{align}
where we adopt $|\Im \tau_i|<\pi$ for $i=1, \dots,N$ on the principal branch of
\textit{BC}$_N$ Harish-Chandra functions and we sum over elements $\mu$ of the
integer cone
$$ Q_+ = \{ \mu = \sum_{a=1}^{N-1} n_a (e_a-e_{a+1}) + n e_N  \, | \, n_a,n \geq 0 \
\textrm{for} \ a=1,\dots, N-1\, \}\ . $$
By inserting this formal expansion into the Calogero-Sutherland eigenvalue equations
one can easily derive equations for the expansion coefficients $\Gamma_\mu$ that may
be solved recursively, at least for generic eigenvalues $\lambda_i$. In a few cases,
explicit formulas for $\Gamma_\mu$ are also known. For $N=2$, for example, the
series expansion of Harish-Chandra functions with generic eigenvalues $\lambda_i$ was
recently worked out in \cite{Isachenkov:2017qgn}, generalizing earlier expressions by
Dolan and Osborn that were only valid for cases in which $\lambda_1-\lambda_2-k_3$
is non-negative integer. The procedure that was employed in \cite{Isachenkov:2017qgn} can in
principle be extended to $N > 2$. This remains an interesting challenge for
future work.

In Heckman-Opdam theory many properties of the Harish-Chandra functions have been obtained
without knowing the explicit series expansions. In particular let us mention that the functions
$\exp(\langle-\lambda+\rho(k),\tau\rangle)\Phi(\lambda;k;\tau)$ are known to be entire functions of
the multiplicities $k_i$ and meromorphic functions of asymptotic data $\lambda_i$, for any
fixed choice of $\tau$ in the fundamental domain. They are known to possess simple poles whenever
the set of $\lambda_i$ satisfies one of the following conditions
\begin{equation} \label{eq:poles}
\langle \lambda_\ast, \alpha\rangle =  \frac{s}{2} \langle \alpha,\alpha \rangle
\quad  \textrm{for}  \quad s = 1,2, \dots \quad , \quad \alpha \in \Sigma^+\ .
\end{equation}
For the poles at $\lambda_\ast = \lambda_{\alpha,n}$, the residues are given by
(see e.g. \cite{OpdamDunkl})
\begin{equation} \label{eq:residues}
\text{Res}_{(\alpha,s)} \Phi(\lambda,k;\tau) \sim \Phi(w(\alpha)\lambda_{\alpha,s},k;\tau)\ .
\end{equation}
where $\sim$ indicates that the relation with the Harish-Chandra function on the
right hand side holds only up to a constant factor. The latter is not known in general,
but it can be found from the series expansion as in \cite{Isachenkov:2017qgn} for $N=2$.
The Harish-Chandra function on the right hand side is related to the one on the left
by acting with an element $w(\alpha) \in W_N$ of the Weyl group on the set of
momenta $\lambda_i$, defined in \eqref{momenta-reflections}.
A complete discussion of poles and residues for $N=2$, including non-generic momenta
$\lambda_i$ can be found in \cite{Isachenkov:2017qgn}.

\subsection{Monodromy representation and wave functions}

The scattering states we have discussed in the previous subsection fail to be good
wave functions for the various real slices one may consider. In fact, at infinity
Harish-Chandra function contains a single plain wave. On the other hand, the latter are
not regular at the walls of the scattering problem. Finding true wave functions
requires to impose regularity conditions at the walls and hence forces us to
consider certain linear combinations of the $2^N N!$ Harish-Chandra functions
with given energy $\varepsilon$.

The behavior of all wave functions at the walls is encoded in the monodromy
representation of the fundamental group. As we saw above, the fundamental
group, which in our case has been identified as the affine braid group,
contains one generator $g_i, i=0, \dots, N$ for each of the walls. The
representation of this generator encodes how wave functions behave as we
continue along a curve that surrounds the wall. Note that all walls possess
real co-dimension two since they are defined by one complex linear equation.
The $2^N N!$-dimensional space of Harish-Chandra functions $\Phi(w\lambda,\tau),
w \in W_N$ carries a representation of the monodromy group. The representation
matrices $M_i = M(g_i)$ are explicitly known from the work of Heckman and Opdam,
see \cite{Isachenkov:2017qgn} for explicit formulas. In the special case of $N=2$,
expressions for two of the three monodromy matrices were also worked out
in the conformal field theory literature \cite{Caron-Huot:2017vep}. Let us
stress that these matrices satisfy the relations \eqref{eq:w1}-\eqref{eq:w3}
that are the defining relation of the affine braid group. In addition they
turn out to obey the following set of Hecke relations
\begin{eqnarray} \label{eq:Hecke}
(M_r-1)(M_r-\gamma_r) & = &  0 \quad , \quad \mathrm{where}\  \\[2mm]
\gamma_0 = e^{\pi i (2k_{2}-1)} \ , \ \gamma_i & = & e^{\pi i(2 k_3-1)} \ , \
\gamma_N = e^{\pi i(2k_1+2k_{2}-1)} \ \nonumber
\end{eqnarray}
for $r = 0, \dots, N$ and $i=1, \dots, N-1$. These may be considered as a
deformation of the relations \eqref{eq:w4}. In this
sense, this monodromy representation of the affine braid group is rather close to
being a representation of the affine Weyl group. For generic values of the
multiplicities $k$ and momenta $\lambda$, the monodromy representation of the
affine braid group on Harish-Chandra functions is irreducible. The precise
condition is
\begin{equation} \label{eq:irred}
2 \frac{\langle \lambda , \alpha\rangle}{\langle\alpha,\alpha\rangle}
\not\in \mathbb{Z}
\quad and \quad
2 \frac{\langle \lambda , \alpha\rangle}{\langle\alpha,\alpha\rangle}
+\frac{k_{\alpha/2}}{2} + k_{\alpha} \not\in \mathbb{Z}
\end{equation}
for all elements $\alpha \in \Sigma^+$. When one of these conditions is violated,
the monodromy representation may contain non-trivial subrepresentations.

In terms of these monodromy matrices, regularity of the wave function $\Phi$ at a
wall $\wall_i$ is equivalent to $\Phi$ being an eigenfunction of the corresponding
monodromy matrix $M_i = M(g_i)$ with unit eigenvalue, i.e.\ $\Phi$ is regular along
$\wall_i$ if and only if $M_i \Phi = \Phi$. There exists a very simple prescription
how to build a function $\Phi$ that is analytic at some subset $\wall_{i_1}, \dots,
\wall_{i_r}$ consisting of $r \leq N$ of the $N$ walls that bound $A^+_N$, i.e \
$i_\nu \neq 0$. For each of these walls there is a generator $w_{i_\nu}$ of the Weyl
group and so our set of $r$ walls is associated with a subgroup $V \subset W_N$ of
the Weyl group that is generated by $w_{i_1}, \dots, w_{i_r}$. Given this subgroup
we now define the following superposition of Harish-Chandra functions
\begin{equation}\label{parabolic-sum}
\Phi^V(\lambdan,k;\tau) = \sum_{w \in V} c(w\lambdan,k) \Phi(w\lambdan,k;\tau)
\end{equation}
where the so-called Harish-Chandra c-function reads
\begin{eqnarray} \label{eq:cfunction}
c(\lambdan,k) & = & \frac{\gamma(\lambda,k)}{\gamma(\rho(k),k)} \quad , \quad
\gamma(\lambdan,k) = \prod_{\alpha\in \Sigma^+}  \gamma_{\alpha}(\lambdan,k) \quad , \\[2mm]
& & \gamma_{\alpha}(\lambdan,k)  =
\frac{\Gamma\left(\frac12 k_{\alpha/2}+ \langle \lambdan, \alpha^\vee\rangle \right)}
{\Gamma\left( \frac12 k_{\alpha/2}+k_\alpha + \langle \lambdan, \alpha^\vee\rangle\right)}.
\end{eqnarray}
For future convenience, let us also introduce
\begin{align} \label{eq:gamma-star}
\gamma_{\alpha}^*(\lambdan,k)  =
\frac{\Gamma\left(1- \frac12 k_{\alpha/2}-k_\alpha - \langle \lambdan, \alpha^\vee\rangle\right)}
{\Gamma\left(1-\frac12 k_{\alpha/2}- \langle \lambdan, \alpha^\vee\rangle \right)}.
\end{align}
Any wave function of the form \eqref{parabolic-sum} turns out be be regular at the
walls $\wall_{i_1}, \dots,\wall_{i_r}$. Physical wave functions on the Weyl chamber
$A^+_N$ are obtained when $V = W_N$ is the entire Weyl group, the most well studied
case in the mathematical literature. For this choice of $V$ we end up with one unique
linear combination of Harish-Chandra functions for each Weyl-orbit of $\lambda$. The
functions $F^+_N = \Phi^{W_N}$ are known as Heckman-Opdam hypergeometric function. They
are close cousins of the Lorentzian hypergeometric functions that were introduced in
\cite{Isachenkov:2017qgn}.  The set of true wave functions $F^+_N(\lambda,\tau)$ of
the Calogero-Sutherland model gives rise to an orthonormal basis of functions on the
wedge $A^+_N$. Let us note, however, that, while the functions $F^+_N$ are analytic in
a neighborhood of $A^+_N$, they fail to be analytic at the wall $\wall_0$. Other real
domains whose boundary contains the wall $\wall_0$, are associated with different
subgroups of the affine Weyl group. Which subgroup one has to sum over in order to
obtain an orthonormal basis of wave functions and the precise form of coefficients 
in this sum depend on the chosen domain for the Calogero-Sutherland scattering problem.

\section{Euclidean inversion formula and defect blocks}

After our sketch of the solution theory for Calogero-Sutherland models we are now in a
position to construct conformal partial waves and blocks. In the next subsection we
shall explain how to build the conformal partial waves explicitly in terms of
Harish-Chandra functions. By definition, conformal partial waves are the physical wave
functions on the Euclidean domain, i.e. single valued solutions of the Casimir equation
in Euclidean kinematics. Our analysis provides one with a complete basis of such wave
functions and hence with a Euclidean inversion formula. In the final subsection we shall
also construct and discuss the conformal blocks that were introduced in section 2.3,
thereby completing the main goal of this work.

\subsection{Euclidean hypergeometrics and inversion formulas}

The Heckman-Opdam hypergeometric functions we described briefly in the final paragraph
of the previous section, provide physical wave functions for the domain $A^+_N$. Their
construction is well known in the mathematical literature. To obtain the Euclidean
inversion formula for defects, we are mostly interested in the physical wave functions
for the Euclidean domain $A^E_N$ that was introduced in eq.\ \eqref{eq:xrange}. As far
as we know, there exists no general theory for these functions, but for the specific
example of $N=2$ that is associated to scalar four-point blocks, such wave functions
have been known in the context of conformal field theory for a long time, see e.g.
\cite{Costa:2012cb,Caron-Huot:2017vep} for explicit formulas in the recent literature.
Here we shall generalize these functions to $N \geq 2$ using the characterization that
was proposed in \cite{Isachenkov:2017qgn}.

Before we can characterize the physical wave functions we need to introduce a bit
of notation. In eq.\ \eqref{eq:xrange} we have introduced the domain $A^E_N$. Of course,
there are quite a few walls within $A_N^E$. When we consider the Calogero-Sutherland problem
it is natural to first formulate it in a smaller domain that is bounded by walls but does not
have walls in the interior. Here we shall describe such a small domain $D^E_N$ and then explain
how to glue $A_N^E$ from the small domain $D^E_N$ and some of its images under the action of
the affine Weyl group. In order to do so we first define the simplex $\triangle_{N-1}$ that
is parametrized by an ordered set of $N-1$ angles $\theta_i$
\begin{equation}
\triangle_{N-1} := \{ (\theta_i, \dots, \theta_{N-1})\, | \, \theta_i \in [0, \pi/2 [
; \theta_i \geq \theta_j \ \mbox{for}\  i < j \} \ .
\end{equation}
We can then introduce the domain $D^E_N$ as a semi-infinite cylinder over $\triangle_{N-1}$,
i.e.
\begin{equation}
D_N^E \:= \{ (\vartheta, \theta_i)\, | \, \vartheta \in \mathbb{R}^+_0 \, ;\, (\theta_i)
\in \triangle_{N-1} \, \} \ .
\end{equation}
The hypercubic base of our the Euclidean domain $A^E_N$ that was introduced in eq.\ \eqref{eq:xrange}
can be triangulated into a disjoint union of the simplex $\triangle_{N-1}$ an its reflections under
the following subgroup $W^B_N$ of the Weyl group $W_N$,
\begin{align}\label{eq:W_B}
W^B_N \colonequals \{w_2, \dots, w_{N-1}, w_N  \,\, | \, \text{ relations of } W_N\} \
\subset W_N\,.
\end{align}
More precisely, our Euclidean domain $A^E_N$ can be decomposed as
\begin{equation} \label{eq:AENDEN}
A_N^E \ =\ \bigsqcup_{w\in \varkappa \cdot W^B_N}\,  w D_N^E,
\end{equation}
where $\varkappa$ is an element of affine Weyl group which simultaneously shifts all the
angular variables. Explicitly, $\varkappa$ acts on the coordinates as $\varkappa:
\theta_j \mapsto \theta_j+\pi/2$ for $j=1, \dots,N-1$ or, equivalently, in terms of the
variables $\tau_j$, it is given by  $\varkappa: \tau_{j+1} \mapsto \tau_{j+1}+i\pi$, $j=1,
\dots, N-1$, while leaving $\tau_1$ invariant. Let us stress that in the decomposition
formula \eqref{eq:AENDEN} the Weyl group elements $w$ act on coordinates, not on momenta
as in most other formulas.

The boundary of $D^E_N \subset \mathbb{R}^{N}$ runs along various walls of our
Calogero-Sutherland problem. In fact, the simplex $\triangle_{N-1}$ which appears
at $\tau_1=0$, runs along the wall $\omega_N$ acted upon with the Weyl reflection
$w_1 w_2 \cdots w_{N-1}$. There are also two semi-infinite
cells of the boundary defined by $\tau_N=0$ and $\tau_{2}=i\pi$
which are part of the wall $\omega_N$, and of its image under the Weyl
reflection $w_2 \cdots w_{N-1}$, respectively. Finally, the boundary
components at $\tau_A=\tau_{A+1}, A=2, \dots , N-1$ run along
the walls $\omega_A$ for $A=2, \dots, N-1$.

We are looking for a physical wave function that is regular along the entire boundary
of the domain $D^E_N$. From our description of the boundary in the previous paragraph
it is clear that such a wave function can be characterized through the following set
of monodromy conditions:
\begin{eqnarray}
M_{1}^{-1} \cdots M_{N-1}^{-1} M_N M_{N-1} \cdots M_{1} F^E(\lambda_i;k_a;\tau_i) & = &
F^E(\lambda_i;k_a;\tau_i)\, , \nonumber \\[2mm]
M_N F^E(\lambda_i;k_a;\tau_i) & = &
F^E(\lambda_i;k_a;\tau_i)\, , \label{eq:charF} \\[2mm]
M_A F^E(\lambda_i;k_a;\tau_i)
& = & F^E(\lambda_i;k_a;\tau_i),\, \nonumber
\end{eqnarray}
where $A=2, \dots, N-1$. The conditions we have displayed here do not directly 
impose triviality of the monodromy along $\tau_2 = i \pi$. Note however that the 
monodromy along the wall $\tau_2 = 0$ is given by the matrix $M_{2}^{-1} \cdots 
M_{N-1}^{-1} M_N \cdots M_{2}$. Since the monodromy matrix along this wall is 
simply a product of monodromy matrices we trivialized, the functions $F^E$ 
are automatically regular along $\tau_2 = 0$. According to our discussion above, 
this ensures that the monodromy along the wall $\tau_2 = i \pi$ is trivial as 
well, as long as we impose appropriate discretization conditions on the momenta 
$\lambda$. If the discretization conditions are violated, on the other hand, the 
functions $F^E$ will possess branch cuts along the wall at $\tau_2 = i \pi$. 
\smallskip 

In building the relevant solutions to the set of conditions \eqref{eq:charF}, let us first look at the
case of $N=2$ for which we only need to trivialize the monodromies $M_2$, $M_1^{-1} M_2 M_1$, along with
$\tilde M_2$ which corresponds to the wall $\tau_2=\pi i$.\footnote{$\tilde M_N$ denotes a monodromy matrix corresponding to the wall $\omega_0$, which amounts to taking $M_N=M_N(\lambda; k_a')$ with parameters
$\{k_a'\}=\varrho_1 \circ \tilde \varrho \circ \varrho_1 \{k_a\}$, see \cite{Isachenkov:2017qgn}.} The
corresponding reflections form a Klein-four subgroup
$$ \mathbb{Z}_2^2=\{1,w_2, w_1w_2w_1, w_1w_2w_1w_2\} \subset W_2$$
of our Weyl group $W_2$ for $N=2$. Using the expressions for monodromy matrices from \cite{HeckmanBook},
the solution to eqs.\ \eqref{eq:charF} is seen to take the form
\begin{align} \label{eq:solutionN2}
F^E_{N=2}(\lambda_i;k_a;\tau_i)=\sum_{w\in\mathbb{Z}_2^2} \gamma^{E}_{N=2}(w\lambda,k)
\Phi(w\lambda_i;k_a;\tau_i)\, ,
\end{align}
where
$$\gamma^{E}_{N=2}\left(\lambda, k\right) \colonequals \gamma^*_{e_1-e_2}
\left(\lambda, k\right) \prod_{\alpha \in \Sigma^+ \backslash \{e_1-e_2\}} \gamma_{\alpha}\left(\lambda, k\right)\ . $$
We claim that the functions $F^E$ form a basis of the space of functions on the Euclidean domain provided
that we let $\lambda_i$ run through
$$
\lambda_1 = \frac{d}{4} - \frac{\Delta}{2} = i \frac{\text{p}}{2} \quad \textit{ and } \quad
\lambda_2 = \frac{d-2}{4} + \frac{\ell}{2} \quad \textit{ for } \quad
\ell = 0,2,4, \dots
$$
where $\text{p}$ is a non-negative real number. The simplest way to see that the basis of such
Euclidean hypergeometric functions will be labeled by even spins $\ell$ is to notice that
the monodromy conditions imposed on all non-compact walls of $D_N^E$ are essentially those for
regularity of a $\BC_{1}$ Jacobi polynomial of $\cos \tau_2/i= \cos 2\theta_1$\footnote{By a quadratic
transformation of this Jacobi polynomial, it can be written as a polynomial in $\cos \theta_1$.}
with $\rho_0= d/4-1/2$. The latter is known to form orthogonal system on the 'simplex' $\triangle_1$,
i.e.\ $\tau_2/i\in [0,\pi[$, only for the discrete set of momenta we have displayed.
By symmetry $\tilde \varrho$ of the $\BC_1$ polynomial problem, these uniquely extend to the
eigenfunctions on our base 'hypercube' $\{\theta_1 \in [0, \pi[\}$, preserving the scalar product.
Correspondingly, the Euclidean hypergeometric function above is defined on the whole Euclidean strip
$A_1^E$, starting from the smaller strip $D_1^E$.

With this experience from $N=2$ we now turn to general $N$. The walls of $D_N^E$ whose monodromy
we need to trivialize are in one-to-one correspondence with reflections in the Weyl group. The
latter generate a subgroup $W^E_N$ of the Weyl group $W_N$,
\begin{align} \label{eq:tildeWN}
W^E_N \colonequals \{w_2, \dots, w_{N-1}, w_N, w_E \,\, | \, \text{ relations of } W_N\} \
\subset W_N\,
\end{align}
where we introduced a shorthand
\begin{equation}
w_E \colonequals w_1 w_2 \dots w_N w_{N-1}\dots w_1.
\end{equation}
Let us remind that $D_N^E$ possesses one wall, namely the wall along $\tau_1 = i\pi$ that is not
associated with a reflection. But as we discussed above, its monodromy is trivialized automatically
once we have taken care of all the other walls and imposed the discretization conditions. The subgroup 
$W^E_N$ has index $N$ in $W_N$. To spell out the Euclidean hypergeometric functions in this case, we denote
\begin{align}
\Sigma_{\star}^+\colonequals \{e_1 - e_j \,| \, j=2, \dots , N\},
\end{align}
and
\begin{align}\label{eq:gamma-Eucl}
\gamma^{E}\left(\lambda, k\right) \colonequals \prod_{\alpha\in \Sigma_{\star}^+} \gamma^*_{\alpha}\left(\lambda, k\right) \prod_{\alpha\in \Sigma^+ \backslash \Sigma_{\star}^+} \gamma_{\alpha}\left(\lambda, k\right).
\end{align}
Then the corresponding solution of the monodromy conditions eqs. \eqref{eq:charF} takes the form
\begin{align}\label{eq:solutionN}
F^E(\lambda_i; k_a;\tau_i)=\sum_{w\in W^E_N} \gamma^E(w\lambda, k)
\Phi(w\lambda_i; k_a;\tau_i)\, .
\end{align}
For later use let us note that these functions $F^E$ are invariant under the action of the
Weyl reflection $w_E$, i.e.
$$
F^E(w_E\lambda_i; k_a;\tau_i) = F^E(\lambda_i;k_a;\tau_i) \
$$
simply because the sum over $W^E_N$ includes a sum over $\{1, w_E\}$.
The Euclidean wave function (or partial wave) \eqref{eq:solutionN} is naively a sum over $2^N (N-1)!$
Harish-Chandra (or pure) functions. In fact, though, most of the coefficients vanish once we
impose the appropriate integrality conditions on the eigenvalues $\lambda_i$ (as it
happens in the case of scalar four-point functions), leaving just two non-zero Harish-Chandra
functions with labels $\lambda=(\lambda_1, \lambda_2, \dots , \lambda_N)$ and $w_E\lambda=
(-\lambda_1, \lambda_2, \dots , \lambda_N)$. Namely, we obtain a complete basis
of wave functions if we let $\lambda_i, i=1, \dots, N$ run through the set
\begin{equation} \label{eq:lalq}
\lambda_1 = \frac{d}{4} - \frac{\Delta}{2} = i\frac{\text{p}}{2}  \quad \textit{and } \quad \lambda_{j+1} = \frac{d}{4} + \frac{l_j - j}{2}    \quad \textit{ with } \quad j=1, \dots , N-1, \quad
l_j = 0,2,4, \dots
\end{equation}
where $\text{p}$ is a non-negative real number, as before. Note that the monodromy conditions imposed on
all non-compact walls of $D_N^E$ are essentially those for regularity of a $\BC_{N-1}$ Jacobi polynomial
of $\left(\cos 2\theta_1, \dots \cos 2\theta_{N-1}\right)$\footnote{By a quadratic transformation of this multivariable Jacobi polynomial, it can be written as a polynomial in $\left(\cos \theta_1, \dots , \cos \theta_{N-1}\right)$.} with
$$ \rho_B= \rho_k^{(N-1)} =
(d/4-1/2, d/4-2/2, \dots , d/4-(N-1)/2)\ .
$$
Here, $\rho_k$ is the vector we introduced in eq.\ \eqref{eq:Thetaasym}, but for the
$\BC_{N-1}$ root system. These Jacobi polynomials are known to form an orthogonal system
on a simplex $\triangle_{N-1}$ only if
$(\lambda_2, \dots, \lambda_{N}) \in \rho_B+P^+_B$, where
\begin{align*}
P^+_B=\{(\nu_1, \dots, \nu_{N-1}) \in \mathbb{Z}_{\geq 0}^{N-1}\,\, | \,\, \nu_1 \geq \cdots \geq \nu_{N-1}\}
\end{align*}
is a set of dominant weights of $\BC_{N-1}$ root system.
As in the $N=2$ case, by symmetries of the $\BC_{N-1}$ polynomial problem, these possess a
unique continuation to the eigenfunctions on our base hypercube $\{\theta_j \in [0, \pi[\}$,
$j=1, \dots , N-1$, such that the scalar product is preserved. Correspondingly, the Euclidean
hypergeometric function above is defined on the whole Euclidean domain $A_N^E$, starting
from the smaller domain $D_N^E$. Our basis functions on $A^E_N$ are labeled by Young diagrams
with even row lengths $2\nu_1 \geq \cdots \geq 2\nu_{N-1}\geq 0$, $\nu_i\in\mathbb{Z}$,
corresponding to spins $l_i=2\nu_i$, $i=1,\dots , N-1$ of a defect partial wave.

A more formal proof of the orthogonality statement goes via Heckman-Opdam shift operators
\cite{Heckman1991, HeckmanBook}\footnote{See \cite{alg-structures} for a review in the context
of conformal field theory.} as follows. First one writes down the inversion  for $k_3=0$, when
 orthogonality trivially splits into applications of polynomial $\BC_{N-1}$ and non-polynomial
 Jacobi (i.e. $\BC_1$) inversion formulas. One then inserts a resolution of the identity $1=
 G_{-}^{k_3} * G_{+}^{k_3}$ via multiplicity shift operators $G_{\pm}$ for the $k_3$ orbit
 (appropriately normalized on Euclidean wave functions by Harish-Chandra isomorphism
 \cite{alg-structures}) into the scalar product of the $k_3=0$ Euclidean hypergeometric
 functions which, by transposition, gives the result for a countable set of values $k_3=0,1,2,
 \dots$. To finalize, one should apply an analytical argument in the spirit of Carlson's
 lemma \cite{AAR} and continue to a dense subset of multiplicities, see \cite{HeckmanBook}
 for samples of such calculations for Calogero-Sutherland wave functions.
\medskip

As we have just established, the functions we have constructed in eqs.\ \eqref{eq:solutionN} and
\eqref{eq:lalq} form a complete and orthogonal set of wave functions for the Calogero-Sutherland
scattering problem in the Euclidean domain. In particular, we can use them to project correlation
functions $G$ for two defects onto conformal blocks, see also
\cite{HeckmanBook},
\begin{align}\label{eq:Euclideaninversion-tilde}
\mathcal{\tilde  C}(\lambda_i)= \frac{(N-1)!}{(i\pi)^{N-1}} \int_{D^E_N} \prod_{j=1}^{N} d \tau_j\,
 \left|\Theta(\tau_i;k_a)\right|^2 \, F^E(\lambda_i;k_a; \tau_i) \, G(\tau_i)\, .
\end{align}
According to the discussion above, we can extend\footnote{Notice that now we restrict to functions
on the Euclidean region possessing $\BC_{N-1}$ symmetry in the angular variables.} this integral
transform to the whole Euclidean region $A_N^E$, which then reads as
\begin{align} \label{eq:Euclideaninversion}
 \mathcal{C}(\lambda_i)= \int_{A^E_N} d\tau_1\, \frac{\prod_{j=2}^{N}d \tau_j}{\left(2\pi i\right)^{N-1}}\,
 \left|\Theta(\tau_i;k_a)\right|^2 \, F^E(\lambda_i;k_a; \tau_i) \, G(\tau_i)\, .
\end{align}
The measure factor $\Theta$ was introduced in eq.\ \eqref{eq:Thetadef} above and the integration
is over the domain $A^E_N$. Convergence of the above integral is assured if $-1<p<d-1$ for the
setup of two point functions in presence of a defect ($N=2$) and $-1<p-q<3-2N+d$ for the setup
of defect two point functions ($N \geq 2$). In those cases with $N=2$ cross ratios that have
previously appeared in the conformal field theory literature, our normalization differs a bit
from the usual one. We will give precise relations below. For later applications we note that
our conventions guarantee that $\mathcal{C}$ possesses the following shadow symmetry,
\begin{equation}\label{eq:Cshadowsym}
\mathcal{C}\left(\lambda\right)=\mathcal{C}\left(w_E\lambda\right)\, .
\end{equation}
Using the orthogonality properties of the partial waves $F^E$ we can invert formula
\eqref{eq:Euclideaninversion} to decompose the correlation function into a sum/integral
over wave functions,
\begin{align} \label{eq:Euclideaninversionback}
G(\tau_i)=
\sum_{\substack{ l_1 \geq \dots \geq l_{N-1} \geq 0\\ l_i \text{ even} }}^{\infty} \int_0^\infty
\frac{d\text{p}}{4\pi}\,
\mu(\lambda;k_a) \, F^E(\lambda_i; k_a; \tau_i) \,  \mathcal{C}(\lambda_i)\, ,
\end{align}
where $\lambda_i$ are considered as functions of $l_i$ and $p$, see eq.\ \eqref{eq:lalq}, and the measure
$\mu$ is given by
\begin{equation} \label{eq:mudef}
\mu(\lambda;k_a)=\prod_{\alpha\in\Sigma_B^+}\left(\frac{\gamma_{\alpha}
\left(\lambda, k\right)}{\gamma_{\alpha}^*\left(-\lambda,k\right)}\right)
\frac{1}{\gamma^E\left(\lambda, k\right) \gamma^E\left(w_E\lambda, k\right)}\, .
\end{equation}
Here, the product runs over the following a subsystem $\Sigma^+_B$ of the root system $\Sigma^+$,
\begin{equation}
\Sigma^+_B = \{e_i, 2 e_i, e_i\pm e_j|2 \leq i,j
\leq N; i < j \}\ .
\end{equation}
Reflections in $\mathbb{C}^{N-1}$ with respect to the roots of this rank $N-1$ system generate the
Weyl group $W_{N-1}$ of $\BC_{N-1}$. The integral over $\text{p}$ runs along the positive real numbers, or
equivalently the $\Delta$ integration runs along the half-line $\Delta = d/2 - i\text{p}$ of principal series
representations. As usual, if some poles of gamma functions in the measure start to cross this line,
bound states start to appear in the spectrum corresponding to residues of the measure at these poles,
which would be equivalent to a Mellin-Barnes prescription for the corresponding integral in $\lambda_1$ over the full
imaginary line. In particular, one can notice that residues appear for $p>d/2$ in the case of a two point function in presence of a defect ($N=2$) and for $p-q>
2+d/2-N$ in the case of a two point function of defects ($N\geq 2$). If the function $\mathcal{C}(\lambda)|_{\lambda_1=i\text{p}/2}$ has residues in $\text{p}$ to the
bottom of the integration line, a contour should be moreover indented to encircle this residue in such a way that no shadow
contribution is picked\footnote{When pole is exactly on the integration line, a principal value
prescription should be taken.}, in full analogy with the case of four-point function. Using the
shadow symmetry \eqref{eq:Cshadowsym} of the function $\mathcal{C}$, the integration over a half-line
becomes integration over the entire imaginary line, so that by closing contour in the lower half-plane\footnote{As
$\lambda_1=d/4-\Delta/2$, this corresponds to standard conformal field theory convention for residues
in $\Delta$ in the case of a four-point function.} and taking residues with the above prescriptions,
one reproduces a bulk operator product expansion. We conclude the list of subtleties with mentioning
that, if poles of blocks themselves appear in the lower half-plane, they should be taken care of in
order not to mix with physical poles, see our description of poles of Calogero-Sutherland
wave functions in section 4.

Since our formulas for the measure factors $|\Theta|^2$ and $\mu$ in eqs.\  \eqref{eq:Euclideaninversion}
and \eqref{eq:Euclideaninversionback} may look a little abstract at first, let us spell out more explicit
expressions for $N=2$.\footnote{With no loss of generality we choose a setup of two point functions in presence of a defect to write these explicit formulas. The case of a defect two-point function with $N=2$ can be obtained from it by setting $a=0$ and replacing $p \mapsto p-q$.} In this case, eqs. \eqref{eq:Thetadef} and \eqref{eq:mudef} give
\begin{align*}
\frac{\left|\Theta(\tau_i;k_a)\right|^2_{N=2}}{4^{d-1+2a}} = \left(\sinh^2 \vartheta \, \sin^2 \theta_1
\right)^{\frac{d-p}{2}-1}\left( \cosh^2 \vartheta \, \cos^2 \theta_1
\right)^{\frac{p}{2}}\left(\sinh^2 \vartheta +
 \sin^2 \theta_1 \right)^{2a+1}
\end{align*}
and
\begin{align*}
\mu(\lambda; k_a)_{N=2}=&\frac{4^{d-2p-4}}{2\pi}\left(\ell+\frac{d}{2}-1\right)\left(\Delta+\ell-1\right)\left(d-\Delta+\ell-1\right)
\frac{\Gamma\left(\frac{d-p+\ell-1}{2},\frac{d+\ell}{2}-1\right)}{\Gamma\left(\frac{\ell+p+1}{2},\frac{\ell}{2}+1\right)}\\[2mm]
&\times\frac{\Gamma\left(\frac{\Delta-1}{2},
\frac{\Delta-p}{2},\frac{d-\Delta-1}{2}, \frac{d-\Delta-p}{2},\frac{\Delta+\ell}{2}+a,\frac{d-\Delta+\ell}{2}+a\right)}
{\Gamma\left(\pm\left(\Delta-\frac{d}{2}\right),\frac{\Delta+\ell}{2}-a,\frac{d-\Delta+\ell}{2}-a\right)}.
\end{align*}
Here we used the standard notation that a $\Gamma$ function with multiple arguments is given
by a product, i.e.\  $\Gamma(a,X) = \Gamma(a) \Gamma(X)$, and $\Gamma(a\pm b) = \Gamma(a+b)
\Gamma(a-b)$. For higher values of $N$, the inversion formula may be a bit more cumbersome
to write out explicitly, but all necessary formulas were spelled out above. Equation
\eqref{eq:Euclideaninversion} is the Euclidean inversion formula we were after in this
section. It is a vast generalization of the Euclidean inversion formula for scalar four-point
functions.
\medskip

As we have noted above, our normalization conventions for the correlation functions $G$ as well
as for the measure factors differ a bit from those used in the existing conformal field theory
literature on two point functions in the presence of a defect. For a direct comparison one
should apply the following list of re-definitions,
\begin{align}
&\mathcal{F}^{\text{CFT}}\left(\tau_i\right)= 4^{\frac{d}{2}+2a} \, \left(\sinh\frac{\tau_1\pm \tau_2}{4}\right)^{\frac{\Delta_1+\Delta_2}{2}} \left(\cosh\frac{\tau_1\pm \tau_2}{4}\right)^{\frac{\Delta_1+\Delta_2}{2}+2a} \,  G\left(\tau_i\right)\nonumber\\[2mm]
&c^{\text{CFT}}\left(\lambda_i\right)=4^{2\lambda_1}  \,
\gamma^E_{N=2}\left(\lambda,k\right)\, \mathcal{C}\left(\lambda_i\right)\\[2mm]
& F^E_{\text{CFT}}(\lambda_i;k_a;\tau_i)=
\frac{4^{\frac{d-1}{2}+a-2\lambda_1}}{\gamma^E_{N=2}\left(\lambda,k\right)}
 \,   \sinh^{a} \frac{\tau_1\pm \tau_2}{2}\,  F^E_{N=2}(\lambda_i;k_a;\tau_i),\nonumber\\[2mm]
& \left|\Theta(\tau_i;k_a)\right|^2_{\text{CFT}}=
4^{4\lambda_1-d-4a}  \,   \left(\sinh\frac{\tau_1\pm \tau_2}{4}\right)^{-\frac{\Delta_1+\Delta_2}{2}-a} \left(\cosh\frac{\tau_1\pm \tau_2}{4}\right)^{-\frac{\Delta_1+\Delta_2}{2}-3a} \nonumber\\
&\phantom{\left|\Theta(\tau_i;k_a)\right|^2_{\text{CFT}}=}\qquad \times\, \left|\Theta(\tau_i;k_a)\right|_{N=2}^2\, .\nonumber
\end{align}
It seems natural to extend these relations with $a=0$ to defect two-point functions with
an arbitrary number $N$ of cross ratios as
\begin{align}
&\mathcal{F}^{\text{CFT}}\left(\tau_i\right):=2^{d}  \,
 G\left(\tau_i\right)\nonumber\\[2mm]
&c^{\text{CFT}}\left(\lambda_i\right):=4^{2\lambda_1}
\,\gamma^E\left(\lambda,k\right) \, \mathcal{C}\left(\lambda_i\right)\\[2mm]
& F^E_{\text{CFT}}(\lambda_i;k_a;\tau_i):= \frac{4^{\frac{d-1}{2}-2\lambda_1}}{\gamma^E\left(\lambda,k\right)}
\, F^E(\lambda_i;k_a;\tau_i),\nonumber\\[2mm]
& \left|\Theta(\tau_i;k_a)\right|^2_{\text{CFT}}:=
4^{4\lambda_1-d}  \, \left|\Theta(\tau_i;k_a)\right|^2.\nonumber
\end{align}
We leave it to the reader to rewrite the Euclidean inversion formula \eqref{eq:Euclideaninversion}
and the conformal partial wave decomposition \eqref{eq:Euclideaninversionback} explicitly with these
conventions.

\subsection{Defect blocks}

Our final goal is to construct the blocks that we introduced through the
expansion \eqref{eq:CPWexp} in terms of Harish-Chandra functions. As in the
case of four-point blocks, all we need to do is to decompose the conformal
partial waves we built in the previous subsection into a sum of a block and
its shadow. Once this is done, the conformal partial wave expansion
\eqref{eq:Euclideaninversionback} can be split into two parts. Using the
shadow symmetry \eqref{eq:Cshadowsym} of the structure function $\mathcal{C}$
we can use the part containing the shadow block to extend the $p$ integration
in the part with the block to the entire real line, see our discussion after
eq.\ \eqref{eq:Euclideaninversionback} for a bit more details. Through
a contour deformation we obtain the expansion of the correlation function in
terms of conformal blocks, as usual.

In order to construct the desired blocks, let us go back to a subgroup
$W^B_N$ of the Weyl group $W_N$ defined in \eqref{eq:W_B}.\footnote{In the
previous section we briefly considered the action of $W^B_N$ on coordinates
of the  Calogero-Sutherland problem. To avoid confusion let us stress that
here we think of $W^B_N$ as acting on the space of momenta $\lambda_i$.}
Obviously, $W^B_N$ is also a subgroup of $W^E_N$, i.e.\ of the group we
averaged over when we constructed the partial waves. In fact, $W^E_N$
contains just one additional reflection, namely $w_E$ that is not included
in $W^B_N$. From the relations \eqref{eq:w1}-\eqref{eq:w3} we infer immediately
that $w_E$ commutes with all elements of $W^B_N$. Hence, as a set $W^E_N$ can be
decomposed as $W^E_N = W^B_N \cup w_E W^B_N$. Consequently, the Euclidean partial
wave $F^E$ that was defined in eq.\ \eqref{eq:solutionN} may be written as a sum
$$
F^E(\lambda_i;k_a;\tau_i) = F^B(\lambda_i;k_a;\tau_i)+F^B(w_E\lambda_i;k_a;\tau_i)
$$
where $F^B$ is obtained by summing Harish-Chandra functions over the subgroup
$W^B_N$,
\begin{align} \label{eq:block}
F^B(\lambda_i;k_a;\tau_i) = \sum_{w\in W^B_N} \gamma^E(w\lambda,k)
\Phi(w\lambda; k_a;\tau_1, \dots , \tau_N)\ .
\end{align}
If we take care of all prefactors and gauge transformations, we arrive at the following
expressions for the blocks we introduced through the decomposition (\ref{eq:CPWexp}),
\begin{equation} \label{eq:defectblockfinal}
\cblock{f_D}{p,q,d}{\Delta_k,\ell_k}{\vartheta, \theta_i} =  \frac{4^{\frac{d}{2}-2\lambda_1}}
{\gamma^E(\lambda, k)} \cdot F^B(\lambda_i;k_a;\tau_i)
\end{equation}
where the multiplicities $k_a$ on the right hand side are related to the parameters $p,q,d$ on the
left through eq.\ \eqref{eq:Caspqpar}. Moreover, the Calogero-Sutherland momenta $\lambda_i$ on the
right hand side are determined by the conformal weight $\Delta$ and the spin $\ell = (l_1, \dots,
l_{N-1})$ of the intermediate channel of the defect block as
\begin{align}
\lambda_1=\frac{d}{4}-\frac{\Delta}{2}\ \quad \quad
\lambda_{j+1}=\frac{d}{4}+\frac{l_{j}-j}{2}, \,\,   j=1, \dots N-1 \ .
\end{align}
Formulas \eqref{eq:block} and \eqref{eq:defectblockfinal} describe conformal blocks
for configurations of two defects as a linear combination of $2^{N-1} (N-1)!$ Harish-Chandra
functions. All coefficients are given explicitly in eq.\  \eqref{eq:gamma-Eucl}. This
extends the construction of four-point blocks from pure functions that was spelled out in
\cite{Caron-Huot:2017vep} to an arbitrary number $N$ of cross ratios.

In the case $q=0$, the blocks can contain an additional parameter $a$ that also enters
the normalization. Here we will adopt the following normalization
\begin{equation}\label{eq:N2blockfinal}
\cblock{f}{p,a,d}{\Delta,\ell}{x,\bar{x}} =
\frac{4^{\frac{d}{2}+a-2\lambda_1}} {\gamma^E(\lambda, k) } \cdot   \sinh^{a} \frac{\tau_1\pm\tau_2}{2}\,
F^B_{N=2}(\lambda_i;k_a;\tau_i)
\end{equation}
which reduces to eq.\ \eqref{eq:defectblockfinal} with $q=0$ when $a=0$, and behaves as
\begin{align}
	\cblock{f}{p,a,d}{\Delta,\ell}{x,\bar{x}} &\stackrel{x\rightarrow 1, \bar{x}\rightarrow 1}
{\longrightarrow}
\left[(1-x) (1-\bar{x})\right]^{\frac{\Delta-\ell}{2}} \left(2-x-\bar{x}\right)^l\, . 
\end{align}
Hence, our conventions match those in the literature. Note, however, that our
normalization differs from those in \cite{Billo:2016cpy}. In order to obtain their blocks
one has to multiply our blocks by a factor $2^{-\ell}$. Formulas \eqref{eq:block} and
\eqref{eq:N2blockfinal} provide an explicit construction of blocks for the bulk channel
of configurations with $q=0$, i.e. when we deal with two local fields in the presence
of a defect of dimension $p < d-1$. In section 3 we described a few cases in which
such blocks can be obtained through the relation with scalar four-point blocks. The
results of section 5, derived through the solution theory of Calogero-Sutherland models,
do not use this connection to four-point blocks. See, however, our discussion
of another class of such formulas in Appendix B.

\section{Conclusions and outlook}

In this work we developed a systematic theory of conformal blocks for a pair of defects
in a $d$-dimensional Euclidean space. By extending the harmonic analysis approach that
was initiated in \cite{Schomerus:2016epl,Schomerus:2017eny} we were able to derive the
associated Casimir equations systematically. These were shown to take the form of an
eigenvalue problem for an $N$-particle Calogero-Sutherland Hamiltonian, generalizing
the observation of \cite{Isachenkov:2016gim} for four-point blocks. We exploited
known symmetries of the Calogero-Sutherland models to obtain a large set of relations
between blocks, of which only a few special cases were known before. Finally, we
gave a lightning review of Heckman-Opdam theory for the Calogero-Sutherland scattering
problem and applied it to the constructions of defect blocks and the Euclidean
inversion formula. The latter generalizes the inversion formula for scalar
four-point blocks in \cite{Dobrev:1977qv}, see also \cite{Costa:2012cb}.

The Euclidean inversion formula for scalar four point blocks was used in
\cite{Caron-Huot:2017vep} to extract the operator product coefficients from (a double
discontinuity of) the Lorentzian correlator. It would be interesting to extend such a formula
to defects, and in particular to correlation functions of two bulk fields in the presence of
a defect. In \cite{Lemos:2017vnx}, a Lorentzian inversion formula was derived for the
\textit{defect channel} of a single defect with two bulk fields, i.e. for $q=0$. This defect
channel inversion formula allowed to extract information on defect operators from the bulk.
Through a Lorentzian inversion formula for the \textit{bulk channel} of the kind described
above it would be possible to go in the other direction, i.e. to infer properties of the bulk
from information on the defect fields. This process could then be iterated.  One way to obtain
the missing Lorentzian inversion formula for (the bulk channel of) defects is to closely follow
the steps in \cite{Caron-Huot:2017vep}. Alternatively, one should also be able to determine the
kernel of the Lorentzian inversion formula algebraically, as explained in \cite{Isachenkov:2017qgn},
starting from our characterization \eqref{eq:charF} of the Euclidean kernel. We will return to this
problem in forthcoming work.

Another interesting direction concerns the extension to spinning blocks, i.e. to non-trivial
representations of the rotation groups $\SO{d-p}$ and $\SO{d-q}$. When $q=0$, these can be
used to expand correlation functions of two fields with spin, such as e.g. the stress tensor,
in the presence of the defect. The harmonic analysis approach that we used in section 3 to
derive our results on the relation with Calogero-Sutherland Hamiltonians was recently extended
to the case of four bulk fields with arbitrary spin \cite{Schomerus:2016epl,Schomerus:2017eny},
i.e. of $p=0=q$, see also \cite{Feher:2009wp}. It is rather straightforward to include defects
into such an analysis. Going through the relevant group theory, one can see that the stabilizer
subgroup of any given point on the double coset is given by $B = \SO{p-q} \times \SO{|d-p-q-2|}$
which is non-trivial unless the two defects possess the same dimension $p=q$ and $d = 2p+2$.
Consequently, the analysis of spinning defect blocks is similar to the cases studied in
\cite{Schomerus:2017eny}. In any case, the corresponding Casimir equations will take the form of
Calogero-Sutherland eigenvalue equations with a matrix valued potential. It should be rewarding
to work these out, at least in a few examples.

As we mentioned in the introduction, extensions of the conformal bootstrap programme including
correlation functions of two bulk fields in the presence of a defect, have played some role
already both for $d=2$ and higher dimensions. Constraint equations on dynamical data
of the theory arise from the comparison of the two different channels that exist for $q=0$, the
bulk and the defect channel. While the defect channel is entirely determined by the expansion
of bulk fields near the defect, the bulk channel also contains information about the bulk
operator product expansions. It is a relevant challenge to compute dynamical data for defect
two-point functions and to formulate appropriate consistency conditions these quantities need
to satisfy. In this context it might also be interesting to include correlators in non-trivial geometries \cite{Nakayama:2016cim} and at finite temperature \cite{Iliesiu:2018fao,Gobeil:2018fzy,Petkou:2018ynm}.

Let us finally stress, that the Heckman-Opdam theory we sketched in section 4 is only a very
small part of what is known about Calogero-Sutherland models. In fact, the most remarkable
property of the Calogero-Sutherland model is its (super-)integrability. It furnishes a wealth
of additional and very powerful algebraic structure. So far, the only algebra we have seen
above was the Hecke algebra that appeared in the context of the monodromy representation. It
acts in the $2^N N!$-dimensional spaces of Harish-Chandra functions $\Phi(w\lambda;z), w \in
W_N$, i.e.\ in finite dimensional subspaces of functions which all possess the same eigenvalue
of the Hamiltonian. This is just the tip of a true iceberg of algebraic structure that involves
e.g. Ruijsenaars-Schneider models and double affine Hecke algebras, see comments in the
conclusions of \cite{Isachenkov:2017qgn}. We will come back to these an other topics in
forthcoming work.
\bigskip

\noindent
{\bf Acknowledgements:} We want to thank Vsevolod Chestnov, Martina Cornagliotto, Nadav Drukker, Abhijit Gadde, Matthijs Hogervorst, Madalena Lemos, Marco Meineri, and Evgeny Sobko for interesting discussions and comments. This work was 
initiated by a workshop on ``Boundary and Defect Conformal Field Theory: Open Problems and Applications'' in 
October 2017 at Chicheley Hall. We are grateful to the organizers, Matthew Buican and Andrew O'Bannon, for
organizing this meeting. This work was completed while one of us (VS) was visiting the PITP in Vancouver. 
VS is grateful for the support and the warm hospitality of the String Theory group at UBC. MI is supported 
in part by Israel Science Foundation (grant number 1989/14), by the ERC STG grant 335182 and by a Koshland 
Postdoctoral fellowship, partially financed by the Koshland Foundation.

\appendix

\section{Relations between coordinates}

\label{apx:coord}

Let us carry out the steps that we outlined in section 2.2 for a pair of defects of dimension
$p$ and $q$. In embedding space, the location of the $p$-dimensional spherical defect of radius
$R$ is described by the points
\begin{align}
	X_i &= (1,R^2,R e_i) \,, \quad X_{p+2} = (1,R^2,-R e_1) \,, &i=1,\dots, p+1 \,.
\end{align}
Similarly, the tilted $q$-dimensional spherical defect of radius $r$ runs through the following
set of $q+2$ points
\begin{align}
Y_i &= (1,r^2,-r \cos(\theta_i) e_i + r \sin(\theta_i) e_{d-i+1}) \,, &i=1,\dots,q+1\,, \notag\\[2mm]
Y_{q+2} &= (1,r^2,r \cos(\theta_1) e_1 - r \sin(\theta_1) e_d) \,,
\end{align}
where we set $\theta_i=0$ for $i \geq N=\min(d-p,q+2)$. A convenient set of orthonormal vectors $P_\alpha$ and
$Q_\beta$ that are transverse to the two defects, i.e.\ satisfy the conditions $X \cdot P=Y \cdot
Q=0$, is given by
\begin{align}
P_{1} &= \left(\frac{1}{R},-R,\vec{0}\right) \,,\quad P_{i} = (0,0,e_{d-i+2}) \,, &i&=2,\dots,d-p\,,\\[2mm]
Q_{1} &= \left(\frac{1}{r},-r,\vec{0}\right) \,, \notag\\[2mm]
Q_{j} &= (0,0,\sin(\theta_{j-1}) e_{j-1} +  \cos(\theta_{j-1}) e_{d-j+2}) \,, &j&=2,\dots,d-q
 \,.
\end{align}
From these explicit expressions it is easy to compute the matrix $M$ of conformal invariants.
It takes the form
\begin{equation}
M = P^TQ = \left(\begin{array}{@{}ccccc|c@{}}
\cosh \vartheta &              &        &                  &   & \\
                & \cos\theta_1 &        &                  &   & \\
                &              & \ddots &                  &   & \scalebox{1.5}{0} \\
                &              &        & \cos\theta_{N-1} &   & \\
                &              &        &                  & I &
\end{array}\right) \,,
\end{equation}
where $\cosh\vartheta = \frac12\left(\frac{r}{R}+\frac{R}{r}\right)$. We recovered our formula \eqref{eq:matrixM}.

Next we want to determine how the coordinates $x$, $\bar{x}$ in \eqref{eq:xxb} that we
used for configurations with $N=2$ cross-ratios relate to our variables $\vartheta$, $\theta\equiv\theta_1$. The former are defined through two local bulk fields ($q=0$) in presence of a
$p$-dimensional defect. In order to apply eq. \eqref{eq:xxb}, we need to project $Y_1$ and $Y_2$ onto the transverse space, i.e.\ the space spanned by $P_1, \dots, P_{d-p}$:
\begin{align}
	\tilde{Y}_1 &= \left(\frac12\left(1-\frac{r^2}{R^2}\right),\frac12(r^2-R^2),r \sin(\theta) e_d\right) \,,\\
	\tilde{Y}_2 &= \left(\frac12\left(1-\frac{r^2}{R^2}\right),\frac12(r^2-R^2),-r \sin(\theta) e_d\right) \,.
\end{align}
Eq.\ \eqref{eq:xxb} yields
\begin{align}
\frac{(1-x)(1-\bar{x})}{(x \bar{x})^{\frac{1}{2}}} &= -\frac{2Y_1 \cdot Y_2}{(\tilde{Y}_1 \cdot \tilde{Y}_1)^{\frac12}(\tilde{Y}_2 \cdot \tilde{Y}_2)^{\frac12}} = \frac{4}{\sinh^2\vartheta + \sin^2\theta} \,,\\ 
\frac{x+\bar{x}}{2(x \bar{x})^{\frac{1}{2}}}  &= \frac{\tilde{Y}_1 \cdot \tilde{Y}_2}{(\tilde{Y}_1 \cdot \tilde{Y}_1)^{\frac12}(\tilde{Y}_2 \cdot \tilde{Y}_2)^{\frac12}}  = \frac{\sinh^2\vartheta - \sin^2\theta}{\sinh^2\vartheta + \sin^2\theta}\ .
\end{align}
We can solve these two equations for $x$, $\bar{x}$ to obtain the expressions we have anticipated
in eq.\ \eqref{eq:xxbtheta}.
In case of four local operators ($p=q=0$) this construction corresponds to the radial coordinates
\begin{equation}
	\rho = \frac{r}{R}e^{i(\pi-\theta)} = -e^{-(\vartheta+i\theta)} \,,\quad \bar{\rho} = \frac{r}{R}e^{-i(\pi-\theta)} = -e^{-(\vartheta-i\theta)} \,,
\end{equation}
and therefore we get
\begin{equation}
	z = \frac{4\rho}{(1+\rho)^2} = -\sinh^{-2}\frac{\vartheta+i\theta}{2} \equiv 1-x \,,\quad \bar{z} = \frac{4\bar{\rho}}{(1+\bar{\rho})^2} = -\sinh^{-2}\frac{\vartheta-i\theta}{2} \equiv 1-\bar{x} \,.
\end{equation}
This concludes our discussion of relations between cross-ratios.

\section{More relations with scalar four-point blocks} 

In this appendix we want to discuss some formulas that can be used to relate any defect block 
with $N=2$ cross ratios to blocks for scalar four-point function. Let us stress, however, that 
the two relations we are about to discuss involve a continuation of the four-point block beyond 
the Euclidean domain, see discussion below. As we have seen before, a situation with $N=2$ 
cross ratios arises when the dimension $p$ of the first defect is $p = d-2$ and the dimension 
$q$ takes any value $q \leq d-2$. In this case we can relate relevant defect blocks to scalar 
four-point blocks through
\begin{align} \label{eq:duality}
\cblock{f_D}{d-2,q,d}{\Delta,\ell}{x,\bar{x}} & \sim
(-4)^{\frac{\Delta+\ell}{2}}  \left[(1-x)(1-\bar{x})\right]^{\frac{d}{2}-2}
\left(\bar{x}-x\right)^{2-\frac{d}{2}} \notag
\\[2mm] &\hspace*{-2cm} \times \cblock{g}{\frac{d-2q-2}{4},\frac{d-4}{4},3}{\frac{\Delta-\ell}{2}-
\frac{d}{2}+2,
-\frac{\Delta+\ell}{2}}{-\frac{(1-x)(1-\bar{x})}{(\sqrt{x}-\sqrt{\bar{x}})^2},
-\frac{(1-x)(1-\bar{x})}{(\sqrt{x}+\sqrt{\bar{x}})^2}} \ .
\end{align}
Recall that the parameters in the upper row of the argument of $g$ are the parameters $a,b$ and $d$
of the scalar four-point block while the parameters in the lower row are the weight $\Delta$ and the
spin $l$ of the exchanged field. If the pair $(x,\bar x)$ describes a point in the Euclidean domain,
i.e. if $x$ and $\bar x$ are complex conjugate to each other, then cross-ratios in the scalar four-point
block $g$ are real, but not inside the unit interval $[0,1]$:
\begin{equation} 
	z = \sin^{-2}\theta \in [1,\infty) \,,\qquad \bar{z} = -\sinh^{-2}\vartheta \in (-\infty,0) \ .
\end{equation}
This means that the four-point block in the right hand side is neither
in the Euclidean nor in the Lorentzian domain, i.e.\ it is related to the usual four-point block only
through analytic continuation to negative real cross-ratios. Conformal blocks, however, possess branch
cuts along the wall $\omega_{1}$. Since the monodromy along this wall is non-trivial, the result of
the analytic continuation on the path along which we continue from positive to negative real
cross-ratios is not unique. The $\sim$ between the left and the right side is meant to remind us of
this continuation.
Formula \eqref{eq:duality} does correctly encode the match of parameters in the Casimir equations,
though, and the identification of eigenvalues up to the action of the Weyl group. In other words,
the defect block on the left hand side can be written through a linear combination of Harish-Chandra
(or `pure' functions in the terminology of \cite{Caron-Huot:2017vep}) with eigenvalues $\Delta,l$
running through all the images of
\beq
\Delta_g := \frac{\Delta-\ell}{2}-\frac{d}{2}+2 \quad , \quad \ell_g := -\frac{\Delta+\ell}{2} \
\eeq
under the replacements $\ell_g \leftrightarrow 2-d_g-\ell_g$, $\Delta_g \leftrightarrow d_g-\Delta_g$
and $\Delta_g \leftrightarrow 1- \ell_g$  with $d_g = 3$.

A similar discussion applies to the second setup with two cross-ratios, namely when we have two
local operators whose weights differ by $\Delta_{12}=-2a$ in presence of a $p$-dimensional defect.
In this case one finds that
\begin{align} \label{eq:duality2}
\cblock{f}{p,a,d}{\Delta,\ell}{x,\bar{x}} & \sim (-4)^{\frac{\Delta+\ell}{2}+a} (x\bar{x})^{\frac{a}{2}}
\left[(1-x)(1-\bar{x})\right]^{\frac{d}{2}-a-2} \left(\bar{x}-x\right)^{2-\frac{d}{2}} \notag
\\[2mm] &\hspace*{-2cm} \times \cblock{g}{-\frac{d-2p-2}{4},\frac{d-4}{4},3+2a}{\frac{\Delta-\ell}{2}-
\frac{d}{2}+a+2,-\frac{\Delta+\ell}{2}-a}
{-\frac{(1-x)(1-\bar{x})}{(\sqrt{x}-\sqrt{\bar{x}})^2},-\frac{(1-x)(1-\bar{x})}{(\sqrt{x}+\sqrt{\bar{x}})^2}} \,.
\end{align}
The $\sim$ between the left and the right hand side has the same meaning as in eq.\ \eqref{eq:duality}.
In some sense, our relations \eqref{eq:duality} and \eqref{eq:duality2} extend the relation \eqref{eq:firstrel}
from \cite{Billo:2016cpy}. While the latter applies to the very special case of $p=d-2$ and $a=0$ only, our
relations cover any setup with two cross-ratios. While the relation between the cross-ratios $x, \bar x$ and
the arguments of $g$ is a little different in eq.\ \eqref{eq:firstrel}, one central feature is the same: it
maps the Euclidean domain of the defect correlator to a different domain and hence, the function $g$ on the
right hand side of eq. \eqref{eq:firstrel} should also be interpreted as some linear combination of Harish-Chandra
functions with eigenvalues $\Delta_g = \Delta$ and $\ell_g = \ell$ running over the full orbit of the Weyl group.

\bibliography{./aux/biblio}

\providecommand{\href}[2]{#2}\begingroup\raggedright\begin{thebibliography}{10}

\bibitem{Isachenkov:2016gim}
M.~Isachenkov and V.~Schomerus, {\it {Superintegrability of $d$-dimensional
  Conformal Blocks}},  {\em Phys. Rev. Lett.} {\bf 117} (2016), no.~7 071602,
  [\href{http://arxiv.org/abs/1602.01858}{{\tt arXiv:1602.01858}}].

\bibitem{Cardy:1991tv}
J.~L. Cardy and D.~C. Lewellen, {\it {Bulk and boundary operators in conformal
  field theory}},  {\em Phys.Lett.} {\bf B259} (1991) 274--278.

\bibitem{Runkel:2005qw}
I.~Runkel, J.~Fjelstad, J.~Fuchs, and C.~Schweigert, {\it {Topological and
  conformal field theory as Frobenius algebras}},  {\em Contemp. Math.} {\bf
  431} (2007) 225--248, [\href{http://arxiv.org/abs/math/0512076}{{\tt
  math/0512076}}].

\bibitem{Polyakov:1974gs}
A.~Polyakov, {\it {Nonhamiltonian approach to conformal quantum field theory}},
   {\em Zh.Eksp.Teor.Fiz.} {\bf 66} (1974) 23--42.

\bibitem{Mack:1973cwx}
G.~Mack, {\it {Conformal Invariant Quantum Field Theory}},  {\em J. Phys.
  Colloq.} {\bf 34} (1973), no.~C1 99--106.

\bibitem{Ferrara:1973vz}
S.~Ferrara, A.~F. Grillo, G.~Parisi, and R.~Gatto, {\it {Covariant expansion of
  the conformal four-point function}},  {\em Nucl. Phys.} {\bf B49} (1972)
  77--98. [Erratum: Nucl. Phys.B53,643(1973)].

\bibitem{Dolan:2000ut}
F.~Dolan and H.~Osborn, {\it {Conformal four point functions and the operator
  product expansion}},  {\em Nucl.Phys.} {\bf B599} (2001) 459--496,
  [\href{http://arxiv.org/abs/hep-th/0011040}{{\tt hep-th/0011040}}].

\bibitem{Dolan:2003hv}
F.~Dolan and H.~Osborn, {\it {Conformal partial waves and the operator product
  expansion}},  {\em Nucl.Phys.} {\bf B678} (2004) 491--507,
  [\href{http://arxiv.org/abs/hep-th/0309180}{{\tt hep-th/0309180}}].

\bibitem{Dolan:2011dv}
F.~Dolan and H.~Osborn, {\it {Conformal Partial Waves: Further Mathematical
  Results}},  \href{http://arxiv.org/abs/1108.6194}{{\tt arXiv:1108.6194}}.

\bibitem{Pappadopulo:2012jk}
D.~Pappadopulo, S.~Rychkov, J.~Espin, and R.~Rattazzi, {\it {OPE Convergence in
  Conformal Field Theory}},  {\em Phys.Rev.} {\bf D86} (2012) 105043,
  [\href{http://arxiv.org/abs/1208.6449}{{\tt arXiv:1208.6449}}].

\bibitem{Hogervorst:2013sma}
M.~Hogervorst and S.~Rychkov, {\it {Radial Coordinates for Conformal Blocks}},
  {\em Phys. Rev.} {\bf D87} (2013) 106004,
  [\href{http://arxiv.org/abs/1303.1111}{{\tt arXiv:1303.1111}}].

\bibitem{Hogervorst:2013kva}
M.~Hogervorst, H.~Osborn, and S.~Rychkov, {\it {Diagonal Limit for Conformal
  Blocks in $d$ Dimensions}},  {\em JHEP} {\bf 08} (2013) 014,
  [\href{http://arxiv.org/abs/1305.1321}{{\tt arXiv:1305.1321}}].

\bibitem{Isachenkov:2017qgn}
M.~Isachenkov and V.~Schomerus, {\it {Integrability of Conformal Blocks I:
  Calogero-Sutherland Scattering Theory}},
  \href{http://arxiv.org/abs/1711.06609}{{\tt arXiv:1711.06609}}.

\bibitem{McAvity:1995zd}
D.~McAvity and H.~Osborn, {\it {Conformal field theories near a boundary in
  general dimensions}},  {\em Nucl.Phys.} {\bf B455} (1995) 522--576,
  [\href{http://arxiv.org/abs/cond-mat/9505127}{{\tt cond-mat/9505127}}].

\bibitem{Billo:2016cpy}
M.~Bill\'o, V.~Gon\c{c}alves, E.~Lauria, and M.~Meineri, {\it {Defects in
  conformal field theory}},  {\em JHEP} {\bf 04} (2016) 091,
  [\href{http://arxiv.org/abs/1601.02883}{{\tt arXiv:1601.02883}}].

\bibitem{Lauria:2017wav}
E.~Lauria, M.~Meineri, and E.~Trevisani, {\it {Radial coordinates for defect
  CFTs}},  \href{http://arxiv.org/abs/1712.07668}{{\tt arXiv:1712.07668}}.

\bibitem{Liendo:2016ymz}
P.~Liendo and C.~Meneghelli, {\it {Bootstrap equations for $ \mathcal{N} $ = 4
  SYM with defects}},  {\em JHEP} {\bf 01} (2017) 122,
  [\href{http://arxiv.org/abs/1608.05126}{{\tt arXiv:1608.05126}}].

\bibitem{Guha:2018snh}
S.~Guha and B.~Nagaraj, {\it {Correlators of Mixed Symmetry Operators in Defect
  CFTs}},  \href{http://arxiv.org/abs/1805.12341}{{\tt arXiv:1805.12341}}.

\bibitem{Gaiotto:2013nva}
D.~Gaiotto, D.~Mazac, and M.~F. Paulos, {\it {Bootstrapping the 3d Ising twist
  defect}},  {\em JHEP} {\bf 1403} (2014) 100,
  [\href{http://arxiv.org/abs/1310.5078}{{\tt arXiv:1310.5078}}].

\bibitem{Liendo:2012hy}
P.~Liendo, L.~Rastelli, and B.~C. van Rees, {\it {The Bootstrap Program for
  Boundary $CFT_d$}},  {\em JHEP} {\bf 1307} (2013) 113,
  [\href{http://arxiv.org/abs/1210.4258}{{\tt arXiv:1210.4258}}].

\bibitem{Gliozzi:2015qsa}
F.~Gliozzi, P.~Liendo, M.~Meineri, and A.~Rago, {\it {Boundary and Interface
  CFTs from the Conformal Bootstrap}},  {\em JHEP} {\bf 05} (2015) 036,
  [\href{http://arxiv.org/abs/1502.07217}{{\tt arXiv:1502.07217}}].

\bibitem{Gliozzi:2016cmg}
F.~Gliozzi, {\it {Truncatable bootstrap equations in algebraic form and
  critical surface exponents}},  {\em JHEP} {\bf 10} (2016) 037,
  [\href{http://arxiv.org/abs/1605.04175}{{\tt arXiv:1605.04175}}].

\bibitem{Lemos:2017vnx}
M.~Lemos, P.~Liendo, M.~Meineri, and S.~Sarkar, {\it {Universality at large
  transverse spin in defect CFT}},  \href{http://arxiv.org/abs/1712.08185}{{\tt
  arXiv:1712.08185}}.

\bibitem{Liendo:2018ukf}
P.~Liendo, C.~Meneghelli, and V.~Mitev, {\it {Bootstrapping the half-BPS line
  defect}},  \href{http://arxiv.org/abs/1806.01862}{{\tt arXiv:1806.01862}}.

\bibitem{Rastelli:2017ecj}
L.~Rastelli and X.~Zhou, {\it {The Mellin Formalism for Boundary CFT$_d$}},
  {\em JHEP} {\bf 10} (2017) 146, [\href{http://arxiv.org/abs/1705.05362}{{\tt
  arXiv:1705.05362}}].

\bibitem{Goncalves:2018fwx}
V.~Goncalves and G.~Itsios, {\it {A note on defect Mellin amplitudes}},
  \href{http://arxiv.org/abs/1803.06721}{{\tt arXiv:1803.06721}}.

\bibitem{Hogervorst:2017kbj}
M.~Hogervorst, {\it {Crossing Kernels for Boundary and Crosscap CFTs}},
  \href{http://arxiv.org/abs/1703.08159}{{\tt arXiv:1703.08159}}.

\bibitem{Gadde:2016fbj}
A.~Gadde, {\it {Conformal constraints on defects}},
  \href{http://arxiv.org/abs/1602.06354}{{\tt arXiv:1602.06354}}.

\bibitem{Fukuda:2017cup}
M.~Fukuda, N.~Kobayashi, and T.~Nishioka, {\it {Operator product expansion for
  conformal defects}},  {\em JHEP} {\bf 01} (2018) 013,
  [\href{http://arxiv.org/abs/1710.11165}{{\tt arXiv:1710.11165}}].

\bibitem{Kobayashi:2018okw}
N.~Kobayashi and T.~Nishioka, {\it {Spinning conformal defects}},
  \href{http://arxiv.org/abs/1805.05967}{{\tt arXiv:1805.05967}}.

\bibitem{Calogero:1970nt}
F.~Calogero, {\it {Solution of the one-dimensional N body problems with
  quadratic and/or inversely quadratic pair potentials}},  {\em J. Math. Phys.}
  {\bf 12} (1971) 419--436.

\bibitem{Sutherland:1971ks}
B.~Sutherland, {\it {Exact results for a quantum many body problem in
  one-dimension. 2.}},  {\em Phys. Rev.} {\bf A5} (1972) 1372--1376.

\bibitem{Heckman:1987}
G.~Heckman and E.~Opdam, {\it {Root systems and hypergeometric functions. I}},
  {\em Compositio Mathematica} {\bf 64.3} (1987) 329--352.

\bibitem{Schomerus:2016epl}
V.~Schomerus, E.~Sobko, and M.~Isachenkov, {\it {Harmony of Spinning Conformal
  Blocks}},  {\em JHEP} {\bf 03} (2017) 085,
  [\href{http://arxiv.org/abs/1612.02479}{{\tt arXiv:1612.02479}}].

\bibitem{Costa:2011mg}
M.~S. Costa, J.~Penedones, D.~Poland, and S.~Rychkov, {\it {Spinning Conformal
  Correlators}},  {\em JHEP} {\bf 1111} (2011) 071,
  [\href{http://arxiv.org/abs/1107.3554}{{\tt arXiv:1107.3554}}].

\bibitem{Koornwinder-quadratic}
T.~H. Koornwinder, {\it {Quadratic transformations for orthogonal polynomials
  in one and two variables}},  \href{http://arxiv.org/abs/1512.09294}{{\tt
  arXiv:1512.09294}}.

\bibitem{Rains-Vazirani}
E.~M. Rains and M.~Vazirani, {\it {Quadratic Transformations of {M}acdonald and
  {K}oornwinder Polynomials}},  \href{http://arxiv.org/abs/math/0606204}{{\tt
  math/0606204}}.

\bibitem{Caron-Huot:2017vep}
S.~Caron-Huot, {\it {Analyticity in Spin in Conformal Theories}},  {\em JHEP}
  {\bf 09} (2017) 078, [\href{http://arxiv.org/abs/1703.00278}{{\tt
  arXiv:1703.00278}}].

\bibitem{van1983homotopy}
H.~van~der Lek, {\em The homotopy type of complex hyperplane complements}.
\newblock Ph.D.\ {T}hesis, {N}ijmegen, 1983.

\bibitem{HeckmanBook}
G.~Heckman and H.~Schlichtkrull, {\em Harmonic Analysis and Special Functions
  on Symmetric Spaces}.
\newblock Academic Press, 1994.

\bibitem{OpdamDunkl}
E.~M. Opdam, {\em Part I: Lectures on Dunkl Operators}, vol.~Volume 8 of {\em
  MSJ Memoirs}, pp.~2--62.
\newblock The Mathematical Society of Japan, Tokyo, Japan, 2000.

\bibitem{Costa:2012cb}
M.~S. Costa, V.~Gonçalves, and J.~Penedones, {\it {Conformal Regge theory}},
  {\em JHEP} {\bf 12} (2012) 091, [\href{http://arxiv.org/abs/1209.4355}{{\tt
  arXiv:1209.4355}}].

\bibitem{Heckman1991}
G.~J. Heckman, {\it An elementary approach to the hypergeometric shift
  operators of opdam},  {\em Inventiones mathematicae} {\bf 103} (Dec, 1991)
  341--350.

\bibitem{alg-structures}
T.~Bargheer, M.~Isachenkov, and V.~Schomerus, {\it {Integrability of Conformal
  Blocks II: Algebraic Structures, in preparation}}, .

\bibitem{AAR}
{G.E.~Andrews, R.~Askey and R.~Roy}, {\em Special Functions}.
\newblock Cambridge University Press, 1999.

\bibitem{Schomerus:2017eny}
V.~Schomerus and E.~Sobko, {\it {From Spinning Conformal Blocks to Matrix
  Calogero-Sutherland Models}},  {\em JHEP} {\bf 04} (2018) 052,
  [\href{http://arxiv.org/abs/1711.02022}{{\tt arXiv:1711.02022}}].

\bibitem{Dobrev:1977qv}
V.~K. Dobrev, G.~Mack, V.~B. Petkova, S.~G. Petrova, and I.~T. Todorov, {\it
  {Harmonic Analysis on the n-Dimensional Lorentz Group and Its Application to
  Conformal Quantum Field Theory}},  {\em Lect. Notes Phys.} {\bf 63} (1977)
  1--280.

\bibitem{Feher:2009wp}
L.~Feher and B.~G. Pusztai, {\it {Derivations of the trigonometric BC(n)
  Sutherland model by quantum Hamiltonian reduction}},  {\em Rev. Math. Phys.}
  {\bf 22} (2010) 699--732, [\href{http://arxiv.org/abs/0909.5208}{{\tt
  arXiv:0909.5208}}].

\bibitem{Nakayama:2016cim}
Y.~Nakayama, {\it {Bootstrapping critical Ising model on three-dimensional real
  projective space}},  {\em Phys. Rev. Lett.} {\bf 116} (2016), no.~14 141602,
  [\href{http://arxiv.org/abs/1601.06851}{{\tt arXiv:1601.06851}}].

\bibitem{Iliesiu:2018fao}
L.~Iliesiu, M.~Koloğlu, R.~Mahajan, E.~Perlmutter, and D.~Simmons-Duffin, {\it
  {The Conformal Bootstrap at Finite Temperature}},
  \href{http://arxiv.org/abs/1802.10266}{{\tt arXiv:1802.10266}}.

\bibitem{Gobeil:2018fzy}
Y.~Gobeil, A.~Maloney, G.~S. Ng, and J.-q. Wu, {\it {Thermal Conformal
  Blocks}},  \href{http://arxiv.org/abs/1802.10537}{{\tt arXiv:1802.10537}}.

\bibitem{Petkou:2018ynm}
A.~C. Petkou and A.~Stergiou, {\it {Dynamics of Finite-Temperature CFTs from
  OPE Inversion Formulas}},  \href{http://arxiv.org/abs/1806.02340}{{\tt
  arXiv:1806.02340}}.

\end{thebibliography}\endgroup
\bibliographystyle{./aux/JHEP}

\end{document}